\documentclass[accepted]{uai2021arxiv} 
\pdfoutput=1

\usepackage[british]{babel}

\usepackage{natbib} 
\usepackage{mathtools} 
\usepackage{booktabs} 
\usepackage{tikz} 

\newcommand{\beginsupplement}{%
        \setcounter{table}{0}
        \renewcommand{\thetable}{S\arabic{table}}%
        \setcounter{figure}{0}
        \renewcommand{\thefigure}{S\arabic{figure}}%
     }

\title{Gaussian Process Nowcasting: Application to COVID-19 Mortality Reporting}

\author[1]{\href{mailto:<i.hawryluk19@imperial.ac.uk>?Subject=Your Arxiv 2021 paper}{Iwona~Hawryluk}{}}
\author[2]{Henrique~Hoeltgebaum}
\author[1]{Swapnil~Mishra}
\author[2]{Xenia~Miscouridou}
\author[3]{Ricardo~P~Schnekenberg}
\author[1]{Charles~Whittaker}
\author[1]{Michaela~Vollmer}
\author[2]{Seth~Flaxman}
\author[1,**]{\href{mailto:<s.bhatt@imperial.ac.uk>?Subject=Your Arxiv 2021 paper}{Samir~Bhatt}{}}
\author[1,**]{\href{mailto:<t.mellan@imperial.ac.uk>?Subject=Your Arxiv 2021 paper}{Thomas~A.~Mellan}{}}
\affil[1]{%
    Department of Infectious Disease Epidemiology\\
    School of Public Health\\
    Imperial College London\\
    UK
}
\affil[2]{%
    Department of Mathematics\\
    Imperial College London\\
    UK
}
\affil[3]{
    Nuffield Department of Clinical Neuroscience\\
    University of Oxford\\
    UK
}
\affil[**]{
    Contributed equally
}

\begin{document}
\maketitle

\begin{abstract}
Updating observations of a signal due to the delays in the measurement process is a common problem in signal processing, with prominent examples in a wide range of fields. 
An important example of this problem is the nowcasting of COVID-19 mortality: given a stream of reported counts of daily deaths, can we correct for the delays in reporting to paint an accurate picture of the present, with uncertainty? Without this correction, raw data will often mislead by suggesting an improving situation. We present a flexible approach using a latent Gaussian process that is capable of describing the changing auto-correlation structure present in the reporting time-delay surface. This approach also yields robust estimates of uncertainty for the estimated nowcasted numbers of deaths. We test assumptions in model specification such as the choice of kernel or hyper priors, and evaluate model performance on a challenging real dataset from Brazil. Our experiments show that Gaussian process nowcasting performs favourably against both comparable methods, and against a small sample of expert human predictions. Our approach has substantial practical utility in disease modelling --- by applying our approach to COVID-19 mortality data from Brazil, where reporting delays are large, we can make informative predictions on important epidemiological quantities such as the current effective reproduction number.\\
\end{abstract}

\section{Introduction}\label{sec:Introduction}

In many real-world settings, current observations from a noisy signal can be systematically biased, with these biases only being corrected after subsequent updates create more complete data. Often, these updates occur much later in the future due to data processing or reporting delays. Not accounting for these delays would result in biased predictions, while waiting for updates would result in a lack of timely estimates. The need for timely estimates to predict the present is colloquially known as nowcasting, and its importance has been shown in a wide range of fields such as actuarial science, economics, and epidemiology \citep{kaminsky_ibnr, lawless_ibnr, bastos_nowcasting_surveillance, mcgough2020nowcasting}.

 \begin{figure*}[!hbt]
\begin{centering}
\includegraphics[scale=0.2]{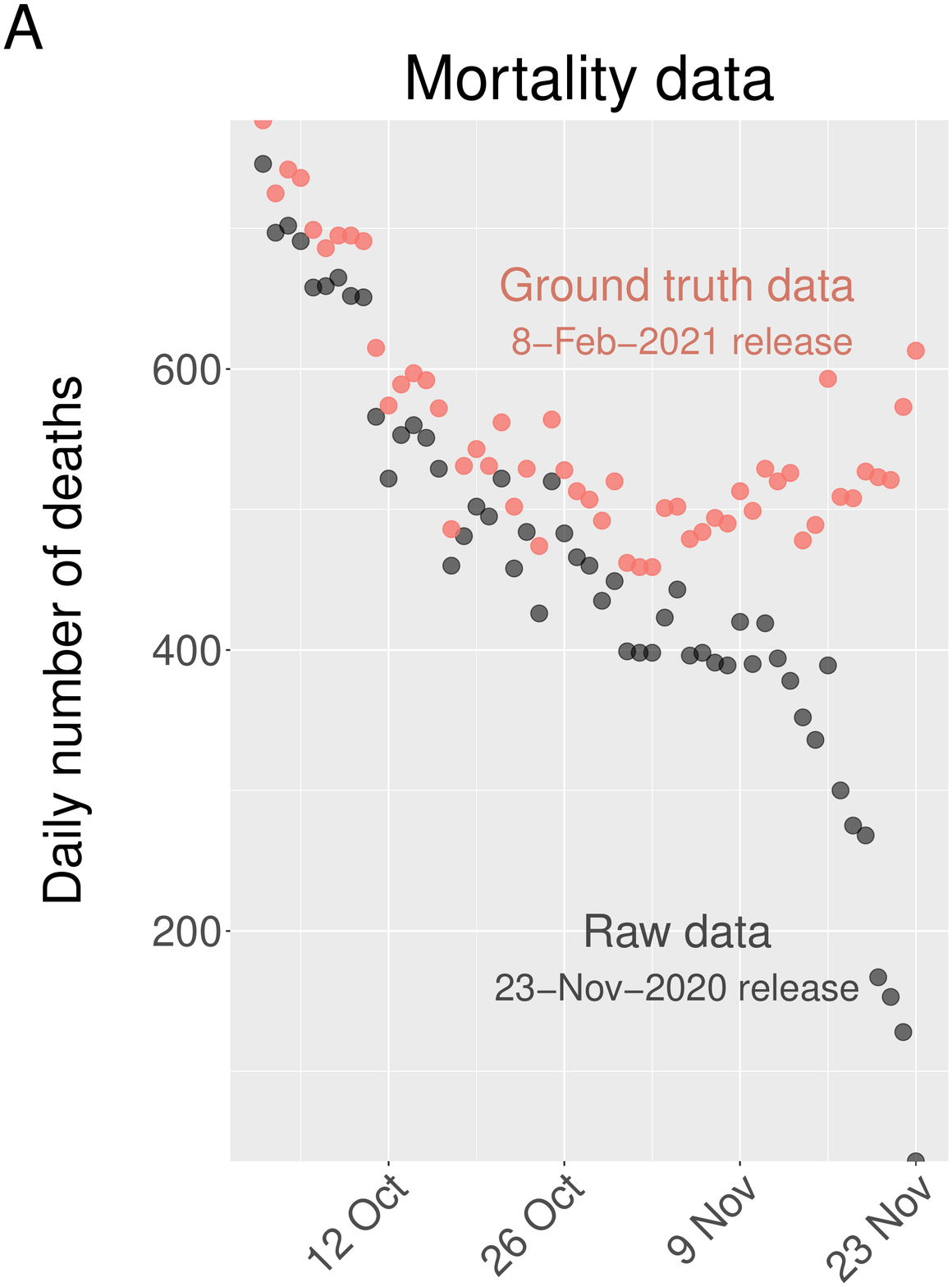}
\includegraphics[scale=0.2]{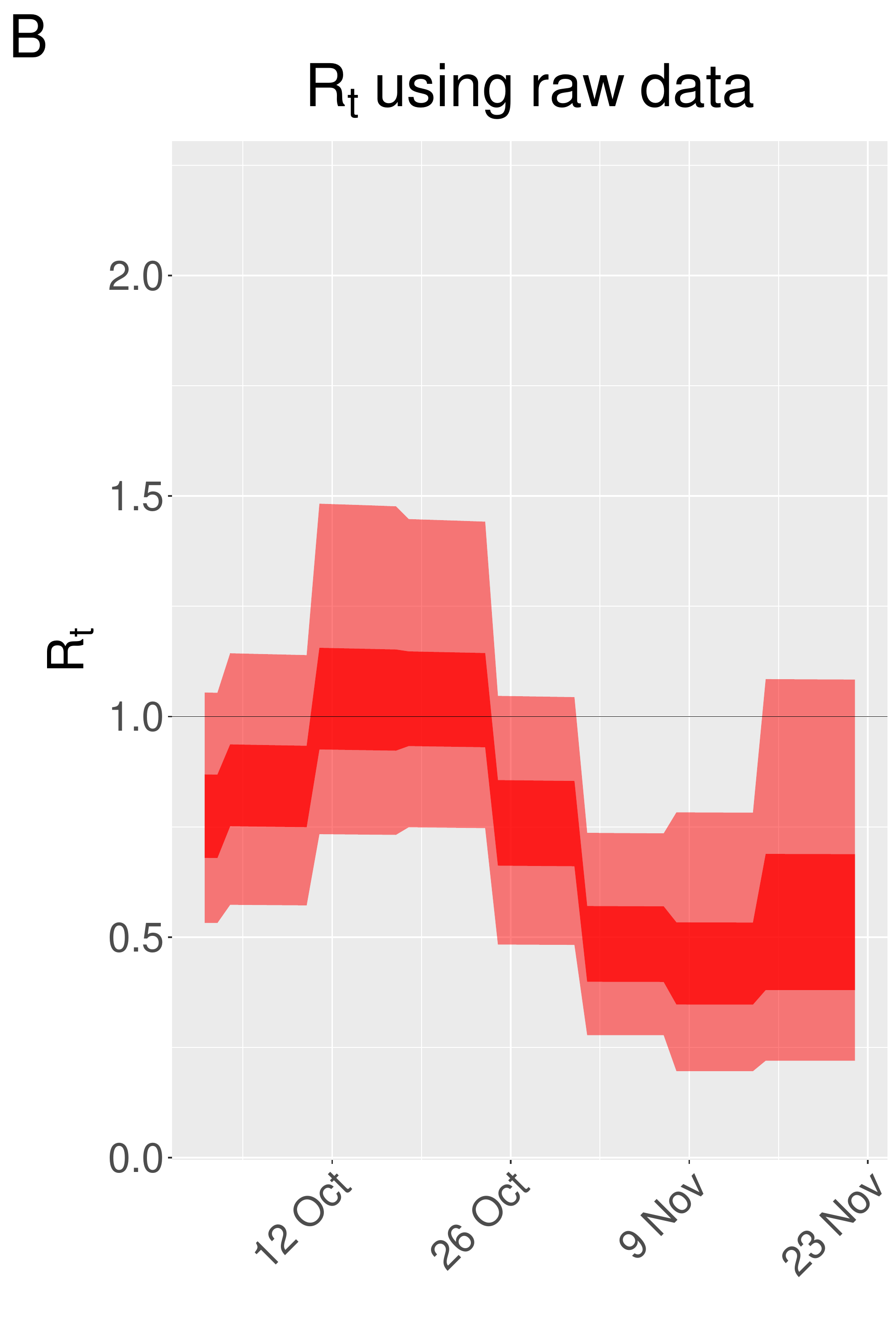}
\includegraphics[scale=0.2]{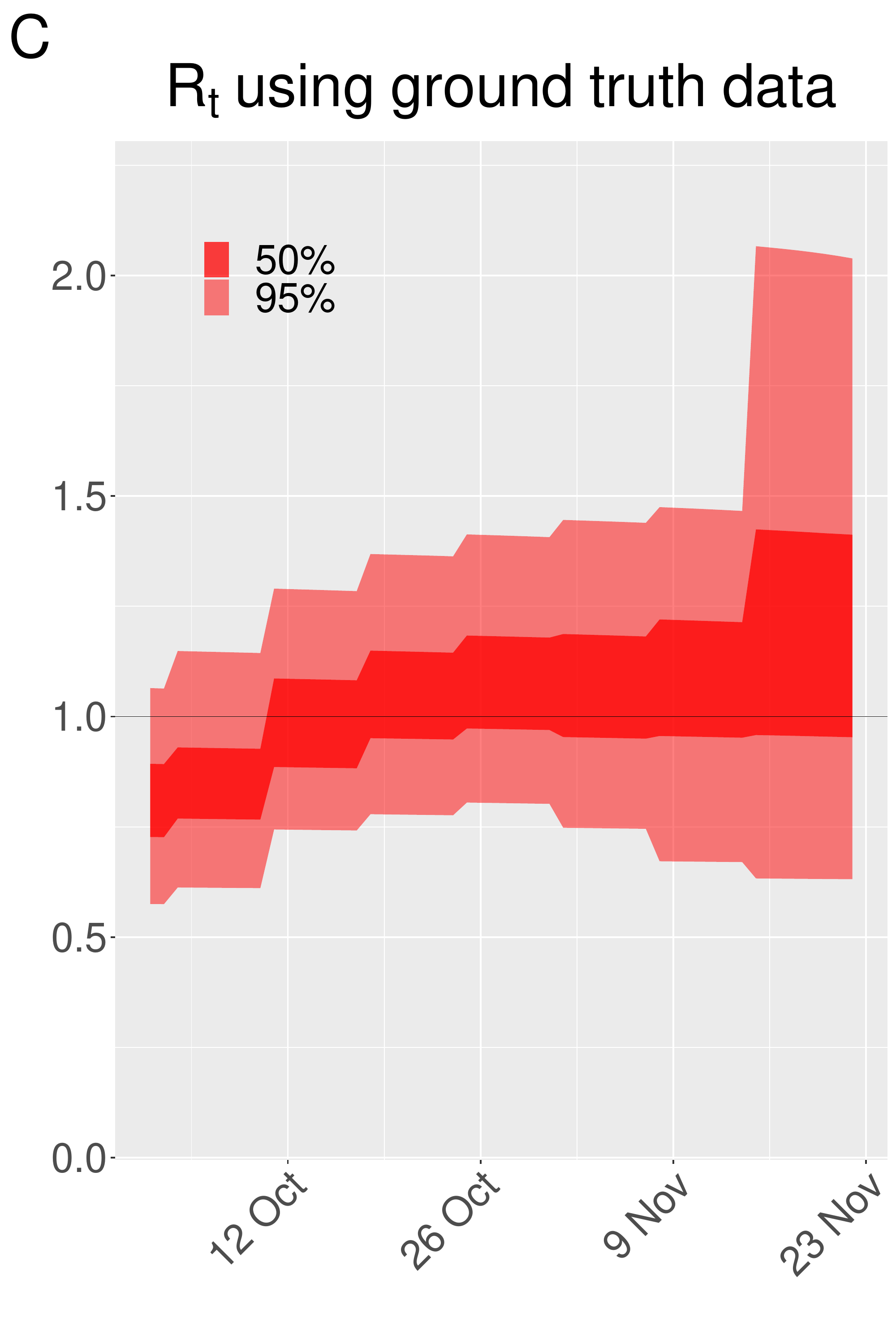}
\includegraphics[scale=0.2]{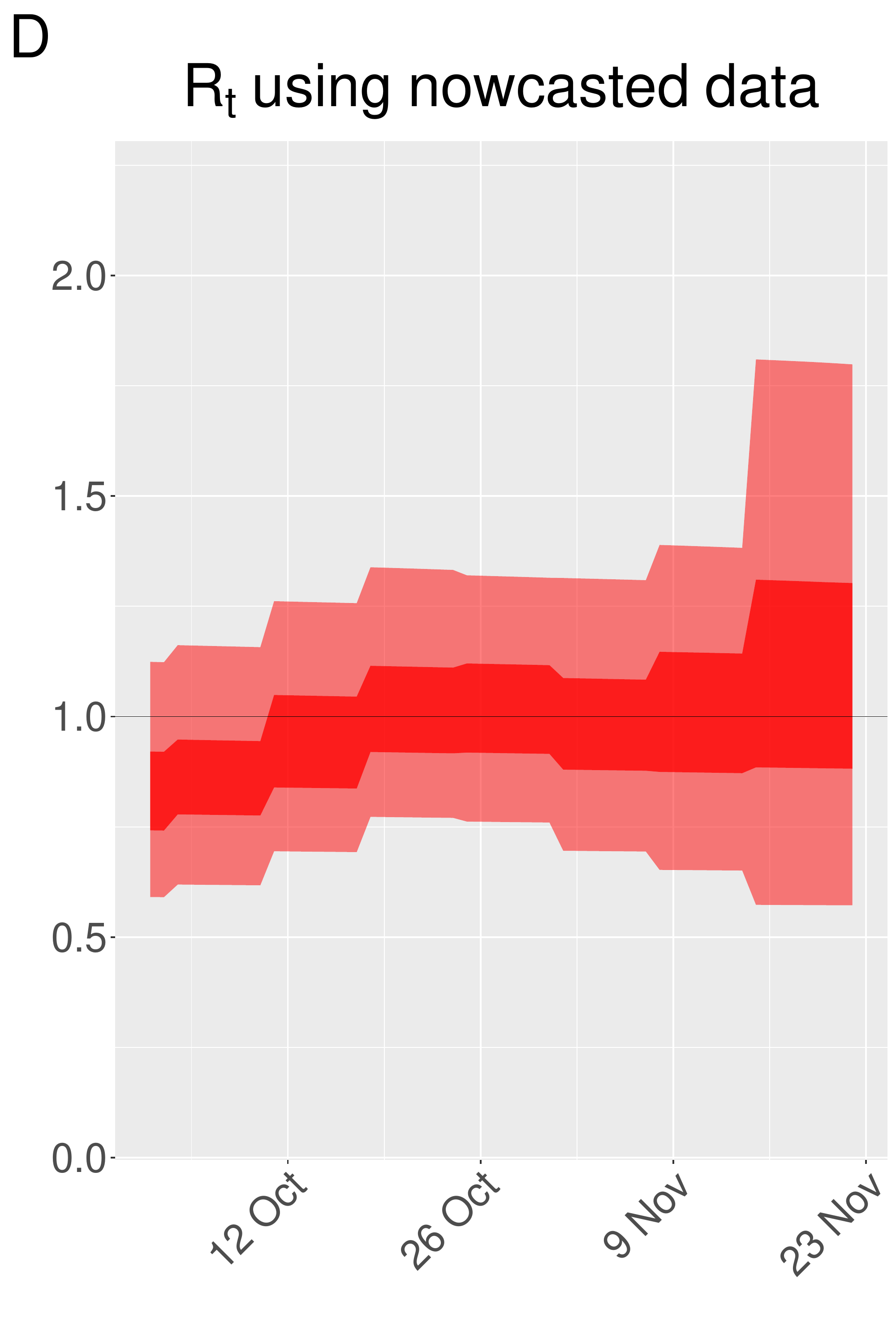}
\par\end{centering}
\caption{A) Reported daily hospital deaths are censored at recent times due to reporting delays. This can be seen by comparing the raw data with a ground truth from two months in the future, when the records have been backdated. B-C) The effective reproduction number $R_t$ for SARS-CoV-2 infections in Brazil from 30-Jun-2020 to 23-Nov-2020, estimated using deaths from the raw reported data released on the 23-Nov-2020, and using a backdated ground truth based on data released on 08-Feb-2021. D) $R_t$ estimates based on nowcasted mortality data. Whereas the raw data results in misleading estimates of $R_t$, with the estimated $R_t < 1$, by applying nowcasting to the deaths counts we achieve a picture of the epidemic closer to the truth.
\label{fig: Rt}}
\end{figure*}

Nowcasting, as defined by \citet{banbura} at the European Central Bank, is the process of predicting the present, the very recent past, and very near future using time series data known to be incomplete. An example from economics is using monthly data to nowcast the current state of important indicators for an economy such as GDP or income. More broadly, nowcasting is relevant for scenarios not only where the data are incomplete, but when the data are comprised of a biased subsample that will be updated in the future retrospectively, following lengthy delays.

In epidemiology, nowcasting is required due to delays in reporting arising from limitations in testing capacity, data curation, and the requirement for pseudonymisation of patient data \citep{bastos_nowcasting_surveillance}. These delays are further compounded by the noise inherent in such data due to limited sampling (typically only a subset of the population is sampled). Throughout this paper, we specifically focus on the delays in the reporting of deaths. An individual dies of a disease on a given day, but the delay between this event and the death being reported (and appearing in the dataset) can be substantial because of the reasons noted above. These reporting delays mask the true \emph{current} state of the epidemic, and have material consequences for our understanding of both present and future evolution of the epidemic. For example, estimation of key epidemiological quantities such as the effective reproduction number ($R_t$) would be systematically biased. Contemporary, real-time and unbiased estimates are necessary for effective public health planning and policy.

In this paper we propose a nowcasting framework based on latent Gaussian processes (GPs). This methodology is used to address the specific problem of delayed reporting in the true incidence of deaths due to COVID-19 in Brazil. 

\subsection{Related Methods}
Previous methods for nowcasting exist in several different contexts. \citet{banbura2014maximum} propose a maximum likelihood approach with a dynamic factor model to predict GDP. \citet{lstm_rainfall_nips} use a deep learning approach based on LSTM to nowcast rainfall intensity. \citet{bastos_infodengue} provide a framework to gather epidemiological information and correct for delays in reporting in Brazilian data. \citet{bastos_nowcasting_surveillance} present a Bayesian hierarchical model for nowcasting applied on data relating to dengue fever and severe acute respiratory infection cases. In \citet{mcgough2020nowcasting} a Bayesian nowcasting approach is proposed that produces accurate estimates that capture the time evolution of the epidemic curve.
Specifically for COVID-19, Bayesian nowcasting approaches have been used to correct for the reporting delays in Bavaria and Sweden \citep{bavaria_covid, sweden_nowcast}. Further discussion around the challenges in estimating reporting delays are also addressed in \citet{nhs_phe_delays}. Finally the problem and background context in Brazil for delays in reporting with corrected data are further explained in \citet{bastos2020covid, villela_RJ}.

Our methods build upon and generalize the NobBS (Nowcasting by Bayesian Smoothing) method originally proposed by \citet{mcgough2020nowcasting}. NobBS is a Bayesian method that produces smooth and accurate nowcasted estimates in the presence of multiple diseases. NobBS allows for both uncertainty in the delay distribution and the evolution of the epidemic curve. While an effective method, NobBS has several limitations, such as inability to pick up fast-occurring changes in the delay distribution, which we overcome in this paper. The extensions we show result in comparable performance for COVID-19 mortality surveillance in Brazil, but present a better fit to the dynamic delays distribution. 
 
\section{Our contributions}
The problem tackled in this paper is conceptually illustrated in Figure~\ref{fig: Rt}. The black points are the data available to us at a given time, and the red the ground truth that is only available much further in the future. It can be seen that the discrepancy between the presently available data and the underlying ground truth data grows markedly as we approach the present -- a distinguishing characteristic of reporting delays. Alongside this, in Figure~\ref{fig: Rt} we also show 3 estimates of the effective reproduction number $R_t$ (defined as the average number of infections an infected individual will go on to infect), obtained using a Bayesian hierarchical renewal-type model \citep{flaxman2020nature, mellan2020report, mishra2020derivation}. Understanding this epidemiological quantity is vital -- $R_t > 1$ results in epidemics growing, while $R_t < 1$ results in epidemics declining. Figure~\ref{fig: Rt}B shows estimates of $R_t$ derived from the raw data, while Figures~\ref{fig: Rt}C and~\ref{fig: Rt}D show estimates of $R_t$ derived from the ground truth data and our nowcasting approach respectively. These plots show that not correcting for delays can lead to a fundamentally different picture of the current epidemic state. Delays in death reporting lead to an underestimation of the true number of deaths in the observed data - the result is a suggestion of a declining epidemic, despite the fact that the epidemic is actually growing.

In this paper we focus on the Brazilian death data from the publicly available hospitalisation database with deaths from both confirmed and suspected COVID-19 diagnostic status \citep{SIVEP}. Our central premise is that using these daily death data alone results in policy decisions being made based on false statistics and trends \citep{villela_RJ}. To facilitate well informed policy making based on unreliable data streams we propose and implement a nowcasting method using latent Gaussian processes. These GPs are capable of capturing the complex correlation structure in delayed data and present an effective means to correct the reporting delays. We use this corrected death data to calculate the effective reproduction number $R_t$ using raw retrospective observed data, nowcasted data and the ground truth updated dataset (Figure~\ref{fig: Rt}).

Our contributions are the following:
\begin{itemize}
    \item We provide a new, flexible and accurate way to correct for delays in reporting. Our framework solves the nowcasting problem through using latent GPs, and provides realistic estimates for the deaths today given incomplete data. Our approach closely predicts the non observed/missing values and simultaneously learns the underlying (latent) data generating mechanisms of the delays. 
    \item We compare our approach to an established alternative method (NobBS), and in a novel contribution, also provide a comparison to a small human expert panel of infectious disease epidemiologists. Domain knowledge is of primary importance for such applications, and is frequently the primary approach taken to interpret data.
    In generating estimates that are improved over both existing computational methods as well as human experts, we demonstrate the utility of our approach.
    \item An important contribution of this work are the results and estimates provided. Implementing our approach enables generating of more accurate estimates of the reproduction number; and in turn, a better understanding of the evolution of the COVID-19 epidemic in Brazil. Our framework is implemented in the easy to use probabilistic program PyStan, and therefore facilitates use in low and middle income settings with limited technical expertise.
\end{itemize}

The structure of the paper is as follows: In section~\ref{sec: Model} we briefly introduce Gaussian processes and describe the latent GP nowcasting models with several variants. In section~\ref{sec: DataAndFits} we describe the data and perform retrospective tests to evaluate the accuracy of the new models and compare them with a sample of human experts predictions. Finally, we discuss the advantages and limitations of the GP nowcasting framework in section~\ref{sec:discussion}.

\begin{figure*}[!h]
\begin{centering}
\includegraphics[scale=0.45]{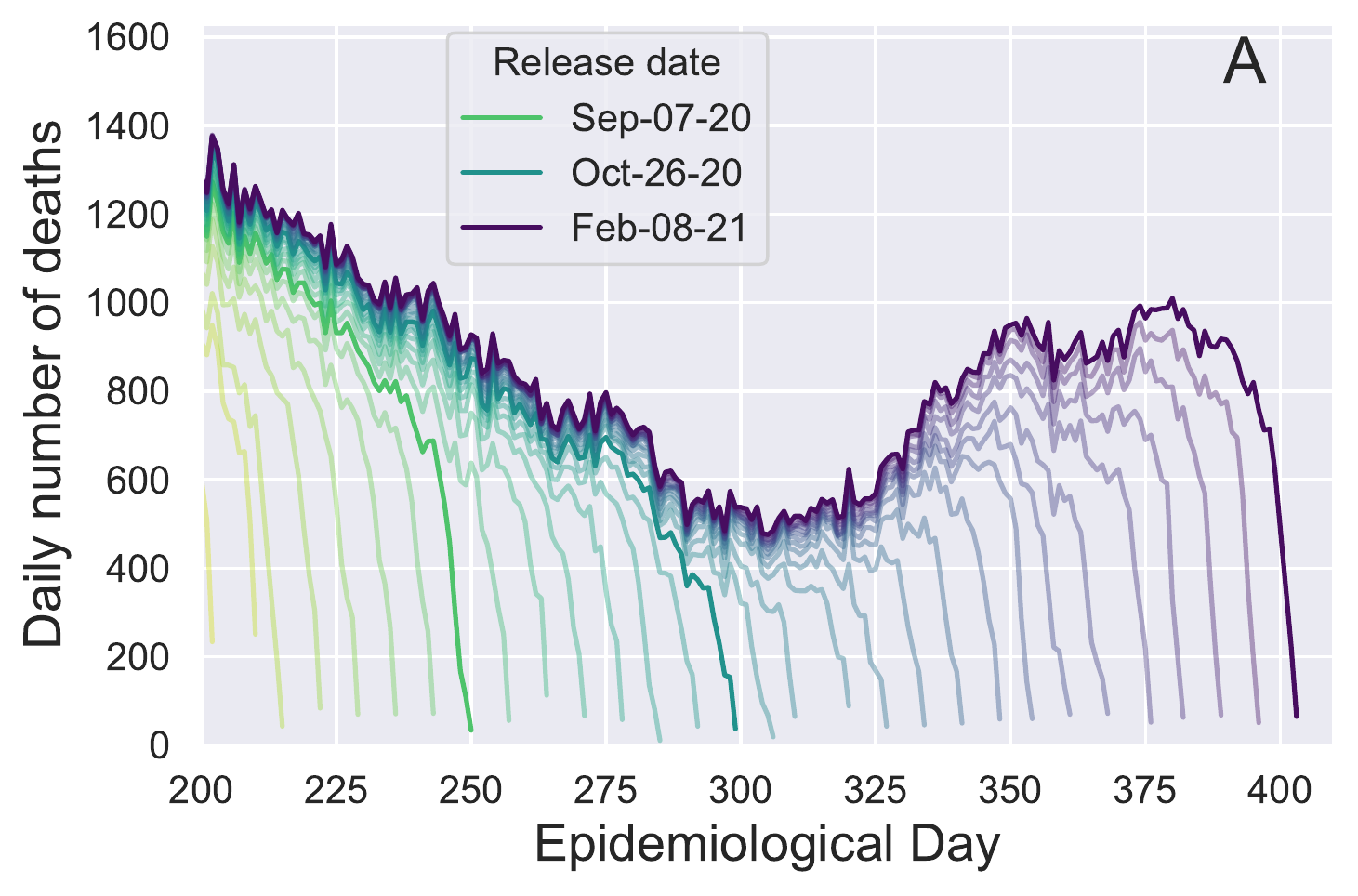}
\includegraphics[scale=0.45]{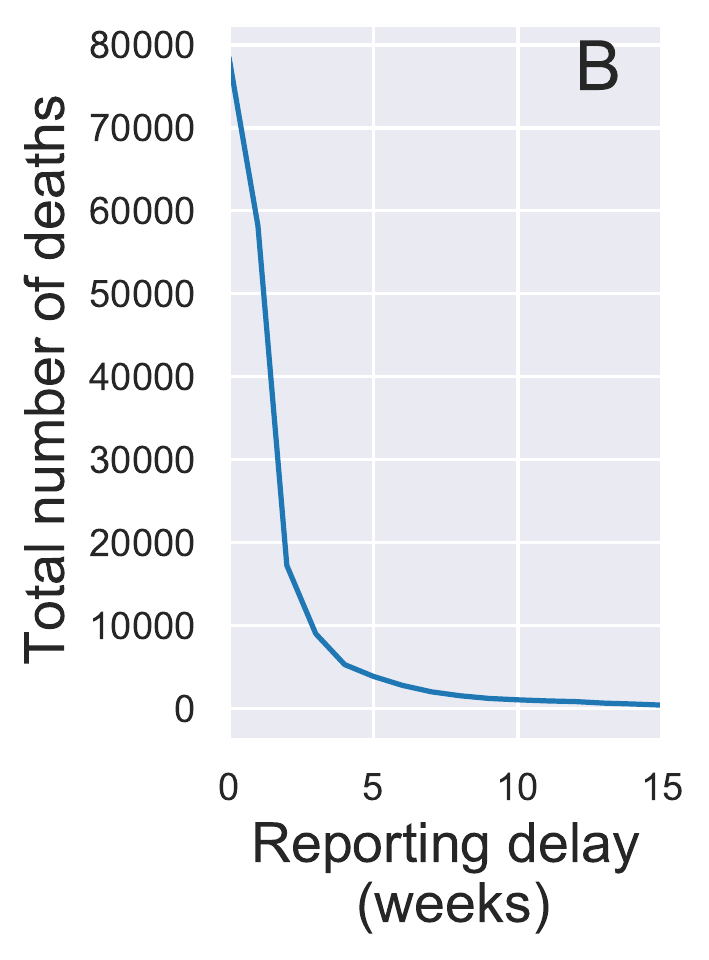}
\includegraphics[scale=0.45]{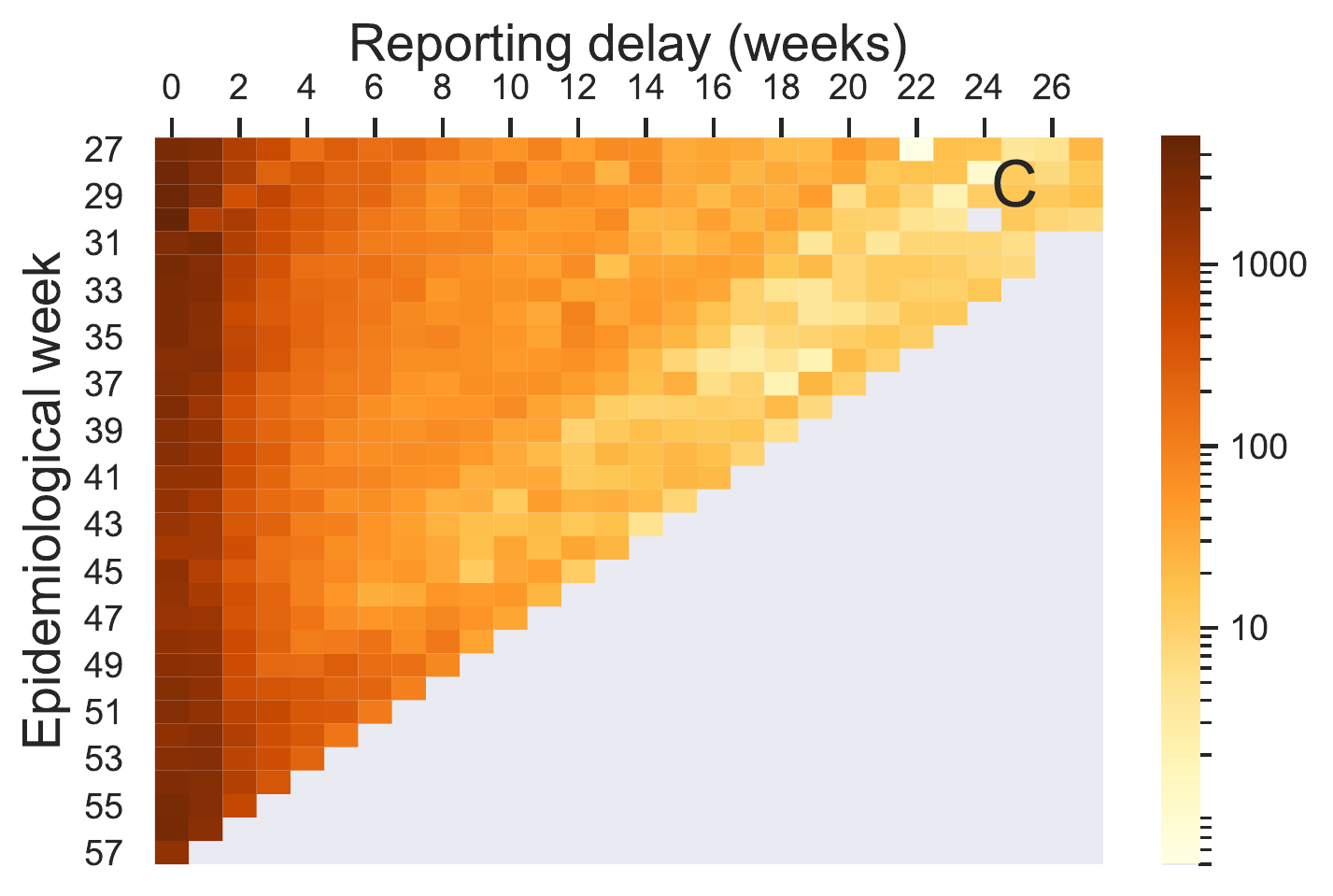}
\par\end{centering}
\caption{A) Daily COVID-19 deaths in Brazil as reported in releases of data between July-2020 and Feb-2021. Each line represents a single release. B) Total number of deaths reported per reporting delay in weeks. Most deaths are reported with delay $\le$ 5 weeks. C) The reporting triangle, showing the number of COVID-19 deaths reported for each week with specific reporting delay. Mortality data were obtained from the SIVEP-Gripe hospitalisation database \citep{SIVEP}. \label{fig: EDA}}
\end{figure*}

\section{Gaussian Process nowcasting}\label{sec: Model}
\subsection{Nowcasting}\label{sec:generic-nowacsting-model}
Let $n_{t}$ denote the response variable of interest that needs to be nowcasted at time $t$. In this paper $n_{t}$ represents the reported COVID-19 mortality in week $t$. The mortality observations, in general, consist of measurements from an online data source, subject to distributed observation delays. The central task of nowcasting approaches is to identify a regular time-delay structure, and use this to estimate $n_{t}$, at a time when it has only been partially observed. The structure that nowcasting identifies is the additive decomposition of the observable over the reporting delay $d$. That is, the true signal at a given time $t$ is the sum over all the delayed partial observations for that time:
\begin{equation}\label{eq: additive_ntd}
n_{t} = \sum_d n_{t,d}\,.
\end{equation}
The intuition behind this formulation is that the "true" deaths that occurred at time $t$ are distributed over various delays $d$ due to the delays in reporting them.

A visual example of partial observation at recent times is the right-censored epidemiological data shown in Figure~\ref{fig: EDA}A. For all data releases, we observe precipitous declines in contemporary data, which is then subsequently revised upwards as the data becomes more complete. In the COVID-19 context this occurs due to time lags in registering and reporting death certificates \citep{villela_RJ}. Figure~\ref{fig: EDA}B shows that most deaths are reported to near completeness after around 5 weeks, and 90\% are reported within 10 weeks. Splitting the data up by delay index we form a 2D array in time and delay, $n_{t,d}$. The filled-in part of $n_{t,d}$, called the \emph{reporting triangle}, is shown in Figure~\ref{fig: EDA}C. The lower triangle part of this 2D array is missing, since at any time $T$ only the number of deaths reported with delay $d \leq T-t$ are known for each epidemiological week $t$. 

The representation of the data by time and delay, rather than time and reporting date, induces a regular structure --- one that is auto-correlated and approximately monotonically decreasing in delay (Figure~\ref{fig: EDA}C). This relatively simple structure makes this problem amenable to statistical modelling. The lower triangle of the $n_{t,d}$ matrix can be predicted with the model, and therefore an estimate of the true signal is available for any time up to the current time by Eqn.~\ref{eq: additive_ntd}, by summing over the delays. This is a common theme from which variations of nowcasting branch out.

To model the discrete positive values of $n_{t,d}$, we can use a Poisson or a negative-binomial likelihood for overdispersed data:
\begin{equation}\label{eq: ntd}
n_{t,d} \sim \text{NB} (\lambda_{t,d}, r)\,.
\end{equation}
In the negative-binomial case, the dispersion parameter $r$ is a hyperparameter that can be learnt or given an informative prior based on the problem. The latter approach is common among the established Bayesian nowcasting methods \citep{bastos_nowcasting_surveillance, bavaria_covid, mcgough2020nowcasting}.
The mean of the negative binomial, $\lambda_{t_d}$, is often modelled as a random walk \citep{bastos_nowcasting_surveillance} or as an auto-regressive process \citep{mcgough2020nowcasting} along the time dimension, that is joint independent with a learnt vector of delays. 
The second approach is taken in the NobBS model \citep{mcgough2020nowcasting}, used in this paper as a benchmark. Specifically, the NobBS model describes $\lambda_{t_d}$ as:
\begin{equation}
    \text{log}(\lambda_{t,d}) = \alpha_t + \text{log}(\beta_d)\,
\end{equation}
where $\alpha_t$ is a latent signal for week $t$ and $\beta_d$ is the probability of reporting with delay $d$. It is worth noting, that with this approach, the distribution of delays $\beta_d$ is fixed throughout the window of analysis.

\begin{figure}[!h]
\begin{centering}
\includegraphics[scale=0.5]{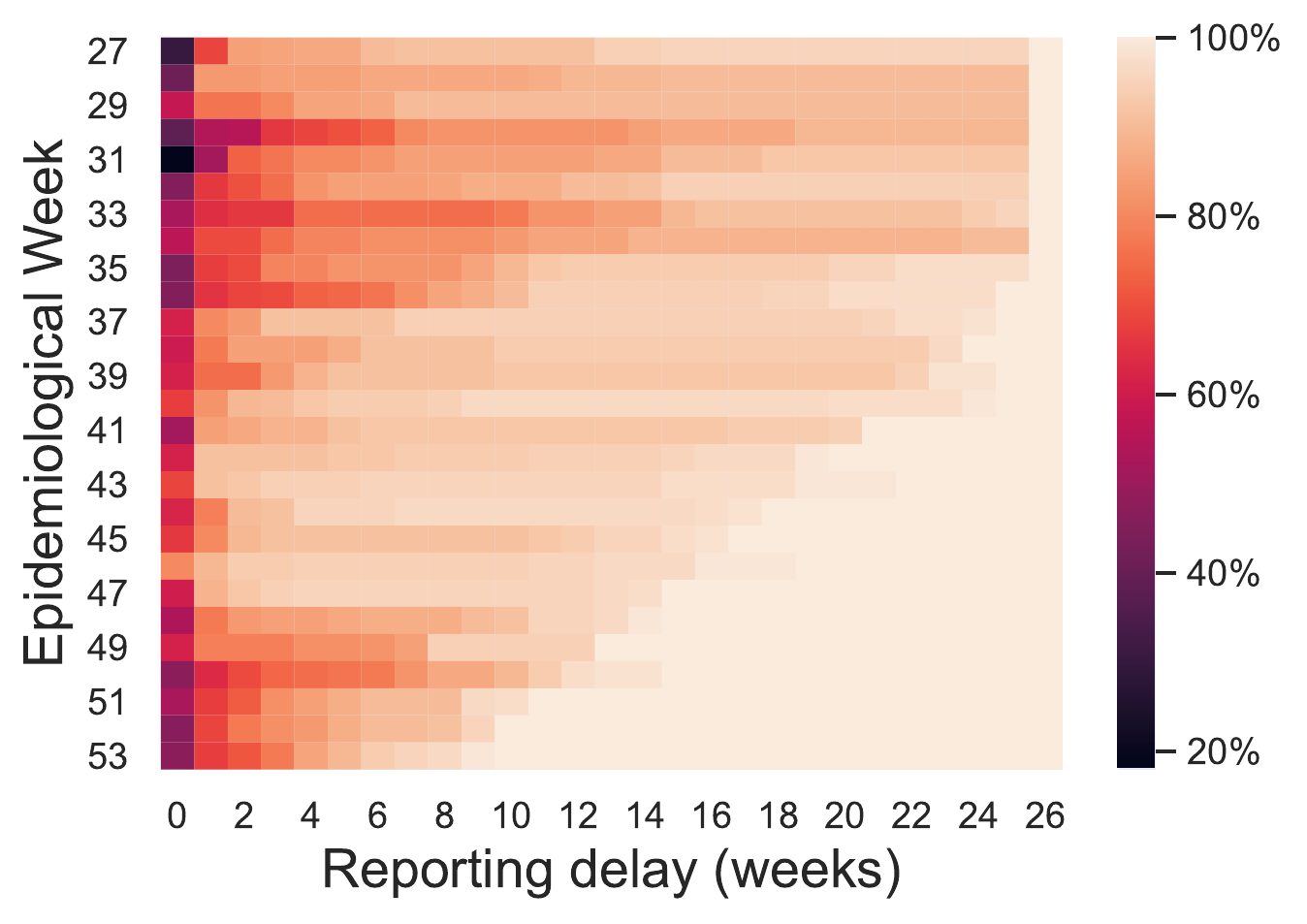}
\par\end{centering}
\caption{The change in the distribution of delays for Manaus (state Amazonas). Each cell in this heatmap shows the percent of all deaths which occurred at week $t$, and were reported with given delay $d$. Here we assume, that up till epidemiological week 53 deaths have been recorded in the database. We use all available SIVEP data up to the release on 5-April-2021. \label{fig: Manaus_delays_changing}}
\end{figure}

This approach has been successful for dengue and influenza surveillance \citep{bastos_infodengue, bastos_nowcasting_surveillance}, but has limitations in terms of the generality of the time-delay covariance structure that can become apparent in more dynamic nowcasting scenarios, such as an evolving epidemic with changing delay distributions (Figure~\ref{fig: Manaus_delays_changing}). Such issues can be minimised by tuning the window over which the static delay vector is estimated, or by manually adding cross-term covariates. Here we employ Gaussian processes as a generic flexible alternative to model arbitrarily structured $\lambda_{t_d}$. The details of this are set out in the following section.
 
\subsection{Latent GP}\label{sec:1D-GP-model}

The introductory model we consider consists of a latent GP with a 1D kernel. In general terms, GPs are a class of Bayesian non-parametric models that define a prior over functions. They are a powerful tool in machine learning, for learning complex functions with applications in both regression and classification problems \citep{rasmussen_gp, wilson2013gpKernelsDiscovery}. In recent years GPs have gained popularity in statistics and in machine learning, due to their flexibility and excellent performance for many spatial and spatiotemporal problems \citep{wilson2013gpKernelsDiscovery, sethFast}, including COVID-19 modelling \citep{neurips_GP_hier}. The covariance function or kernel together with the mean function completely define a GP. The mean function is the base function around which all of the realizations of the GP are distributed. The \emph{covariance kernel} is a crucial component of the Gaussian process, as it describes the covariance of the Gaussian process random variables i.e. how similar two points are. Therefore, the kernel defines the shape of the distribution and which type of functions are more probable. 

One of the most popular choices of covariance kernel, and the one we chose to introduce the model with, is the \emph{squared exponential} kernel, $k_\text{SE}$, with entries defined by a covariance function $k_\text{SE}(\cdot,\cdot)$ such that
\begin{equation}
k_\text{SE}(t_i, t_j) = \alpha^2 \text{exp}\left(-\frac{||t_i - t_j||_2^2}{2\rho^2}\right)\,.    
\end{equation}

The parameter $\alpha$ defines the kernel's variance scale, and $\rho$ is a lengthscale parameter that specifies how nearsighted the correlation between pairs of time points ($t_i$) is. The kernel results in a prior over a set of functions to describe, $\lambda_{t,d}$, the mean of the statistical model in Eqn.~\ref{eq: ntd}. This is modelled as a zero mean log-space latent Gaussian process
\begin{equation}
\text{log}(\lambda_{t,d}) \sim \text{GP}(0, k_\text{SE})\,.
\end{equation}

Due to weak identifiability \citep{rasmussen_gp}, a strategy to identify the hyperparameters $\rho$ and $\alpha$ is to fix the lengthscale $\rho$ to the maximum delay time considered in the nowcasting problem, and learn only the scale parameter $\alpha$. Markov Chain Monte Carlo (MCMC) is used in order to generate posterior summaries for arbitrary (non-normal) latent Gaussian processes.

\subsection{Generalised model}\label{sec:GP-model-gen}
    
\subsubsection{Additive Kernel Model}\label{sec:GP-model-additive}
The basic model introduced above can be extended to provide a more expressive description of the data. The purpose of this is to be able to describe the complex structure in $n_{t,d}$. 

Using the compositional kernel approach \citep{duvenaud_icml, wilson2013gpKernelsDiscovery, wilson2016deep}, we can create a new additive kernel over multiple lengthscales, indexed $s$, as 
\begin{align}
  k_\text{add} &= \sum_s k_s\,,  \nonumber \\
    \text{log}\left(\lambda_{t,d}\right) &\sim \text{GP}(0,k_\text{add})\,.
\end{align}

The lengthscale hyperparameters are fixed or given strongly informative priors, $\rho_s$, while each $\alpha_s$ is learnt.
In the simplest case we consider a kernel with two lengthscale contributions, short- and long-range correlation structure:
\begin{equation}
    k_\text{add}(t_i, t_j) = k_\text{long}(t_i, t_j) + k_\text{short}(t_i, t_j) + \sigma^2\delta_{ij}\,,
\end{equation}  
plus a regularising term with a Kronecker delta function ensuring $\sigma^2$ Gaussian noise is only added when $i = j$. The choice of kernel confers bias that can result in a better generalisation. The logic of this kernel is to split the covariance into two components: (a) a smooth long-range component, used to extrapolate the trend into the unknown part of the reporting triangle where large distances from the observed points exist and (b) a part for describing variation in $n_{t,d}$ over shorter lengthscales. Additionally, the separation of kernels provides a generic method to describe more complex data generating processes -- for example, the long-range kernel can be squared-exponential, while the short-range can be a less smooth type with a different power spectrum such Matérn (1/2). This can be used to create a general statistical model for all of $n_{t,d}$. Furthermore, in this regard the $\delta$ contribution provides a source of regularisation which may be useful if there is reason to believe $n_{t,d}$ values are subject to variation beyond the scope of the basic nowcasting framework. For example, if a death can switch category from a COVID-19 suspected death to a cause other than COVID-19 in later data releases, this could result in a negative $n_{t,d}$ count, which can be modelled as an error to be regularised.

A further modification that can be applied if the time-delay surface $n_{t,d}$ has a complex structure, is to split the data into two components and model each with separate kernels. For example, if delays of 0 or 1 weeks account for a large fraction of total counts, they can be considered separately to delays $>1$. This approach is considered later in section~\ref{subsec: modelFit}. But a more generic formulation is to consider a 2D kernel to fully account for the time-delay correlation structure, which is introduced below.

\subsubsection{2D Kernel Model}\label{sec:GP-model-2D}
As a further expansion of the approach described before, we introduce a separable two dimensional kernel over time and delay, $k((t_i,t_j),(d_i,d_j)) = k_t(t_i,t_j)k_d(d_i,d_j)$. Separable kernels can be efficiently implemented using Kronecker product algebra as described in \citet{sethFast}. 
Specifically, individual Gram matrices for time and delay are combined using the Kronecker product such that
\begin{equation}
  K_{t,d} = K_t \otimes  K_d\,.  
\end{equation}

As before, the kernel can be given an additive structure over multiple lengthscales. For example, 
\begin{align}
    k_{\text{long}}(t,d) &= k_{\text{long}}^t k_{\text{long}}^d\,, \nonumber \\
    k_{\text{short}}(t,d) &= k_{\text{short}}^t k_{\text{short}}^d\,, \nonumber \\
    \text{log}\left(\lambda_{t,d}\right) &\sim \text{GP}(0,k_{\text{long}}(t,d) + k_{\text{short}}(t,d))\,.
\end{align}

This approach captures the relationship between $t$ and $d$. In both 1D and 2D kernel approaches it is possible to perform partial pooling of the models parameters by combining two or more spatial locations with similar features, for example neighbouring states, if limited data is available. In practice however we found that there was a limited gain in doing so as our approach works well with relatively few observations.

\section{Data and Model properties, fit and testing}\label{sec: DataAndFits}
\subsection{Data} \label{sec:data}

The numbers of deaths per date have been extracted from the Brazilian Ministry of Health's Sistema de Informação de Vigilância Epidemiológica da Gripe (SIVEP-Gripe) database \citep{SIVEP}. SIVEP-Gripe is a large publicly available database providing anonymised patient-level records of all individuals who died or were hospitalised with suspected or confirmed COVID-19 in Brazil \citep{bastos2020covid, souza_brazil_sivep, niquini_sari}. New data have been released regularly online, on a weekly basis, in the second half of 2020 considered here. In this study, we extracted all SIVEP-Gripe data releases from 7-July-2020 to 31-May-2021. We consider all cases of suspected or confirmed COVID-19 (class 4 and 5).

There are a number of potential sources of error in the reported SIVEP data. One is underascertainment --- systematic biases which are beyond the scope of correction by this nowcasting methodology. Another source of error is delayed classification. After the initial input of patient's data into the database (usually at the time of hospitalisation), the entry might be later updated with clinical and laboratory data, including confirmatory COVID-19 testing. Further updates will include the outcome and its date (i.e. date of death or date of hospital discharge) and cases receive a final classification. Cases can be classified as confirmed (class 5) or suspected COVID-19 (class 4), or other causes (classes 1-3). Despite being described as a "final classification", reclassification does occur, and is especially common for unknown cases to be reclassified as COVID-19 once results from confirmatory tests are informed to the health authorities. On the other hand, some deaths attributed to suspected SARS-CoV-2 infection are later "removed" from the SIVEP database, due to duplicate filtering or because they are eventually attributed to other diseases. That can cause the number of deaths on certain days to decrease in consecutive data releases, as shown in Figure~\ref{fig: sivep_errors_AM} in the Supplement.

The number of deaths per day as reported by each release is presented in Figure~\ref{fig: EDA}, together with a reporting triangle, showing the distribution of the reporting delays across time. According to the SIVEP-Gripe dataset, over 90\% of all deaths have been reported with delay less than 10 weeks (Figure~\ref{fig: EDA}B). We therefore choose the maximum reporting delay $D$ for our data to be $D=10$, and sum up all deaths which were reported with the delay longer than 10 weeks. Finally, to create the reporting triangle appropriate for our model, we aggregate the data into weeks.

\subsection{Model Fit} \label{subsec: modelFit}
We fit and present 7 models. For 1D kernel GPs, we consider a single SE kernel (1D SE), and additive long- and short-range component kernels (1D SE+SE and 1D SE+Mat). The additive long- and short-range component kernels are also considering splitting the data into across delay greater and less than one (1D SE+SE data-split). Finally we consider a 2D kernel GP model with additive long and short range components (2D additive). The NobBS model of \citet{mcgough2020nowcasting} is fitted and presented for reference of the current state-of-the-art. All models are fitted to the SIVEP-Gripe weekly COVID-19 deaths reported in Brazil, currently available until 31-May-2021.

Posterior samples of the parameters in the models were generated using Hamiltonian Monte Carlo with Stan \citep{nuts, carpenter2017stan}, using the PyStan interface (version 2.19.0.0). For each fit we used 4 chains and 1000 iterations, with 400 iterations dedicated to warm-up. The convergence of each model fit was evaluated by ensuring that $\hat{R} < 1.01$ for each parameter. Traceplots and other MCMC diagnostic measures were also investigated (Supplement, section~\ref{s: diagnostics}).

Each of the models, characterised by the likelihood given in Eq~\eqref{eq: ntd}, and a latent GP part for modelling the $\lambda_{t,d}$ (section~\ref{sec:1D-GP-model}) is trained by supplying the reporting triangle $n_{t,d}$ filled with data available up to the point of the nowcast. Each of the parameters governing the model, such as overdispersion $r$ or lengthscales and variances of the GPs are learnt during the model fit. The best performing hyperparameters of the prior distribution were selected conditioned on the observed results. All of those parameters and their prior densities are given in the Supplement, Table~\ref{tab: priors}. The training of the model and nowcasting through sampling each element of the $n_{t,d}$ matrix is done simultaneously. Specifically, at each iteration parameter values are sampled and immediately used to sample from the negative-binomial distribution to obtain all elements of the $n_{t,d}$ matrix.

\begin{figure}[!h]
\begin{centering}
\includegraphics[scale=0.5]{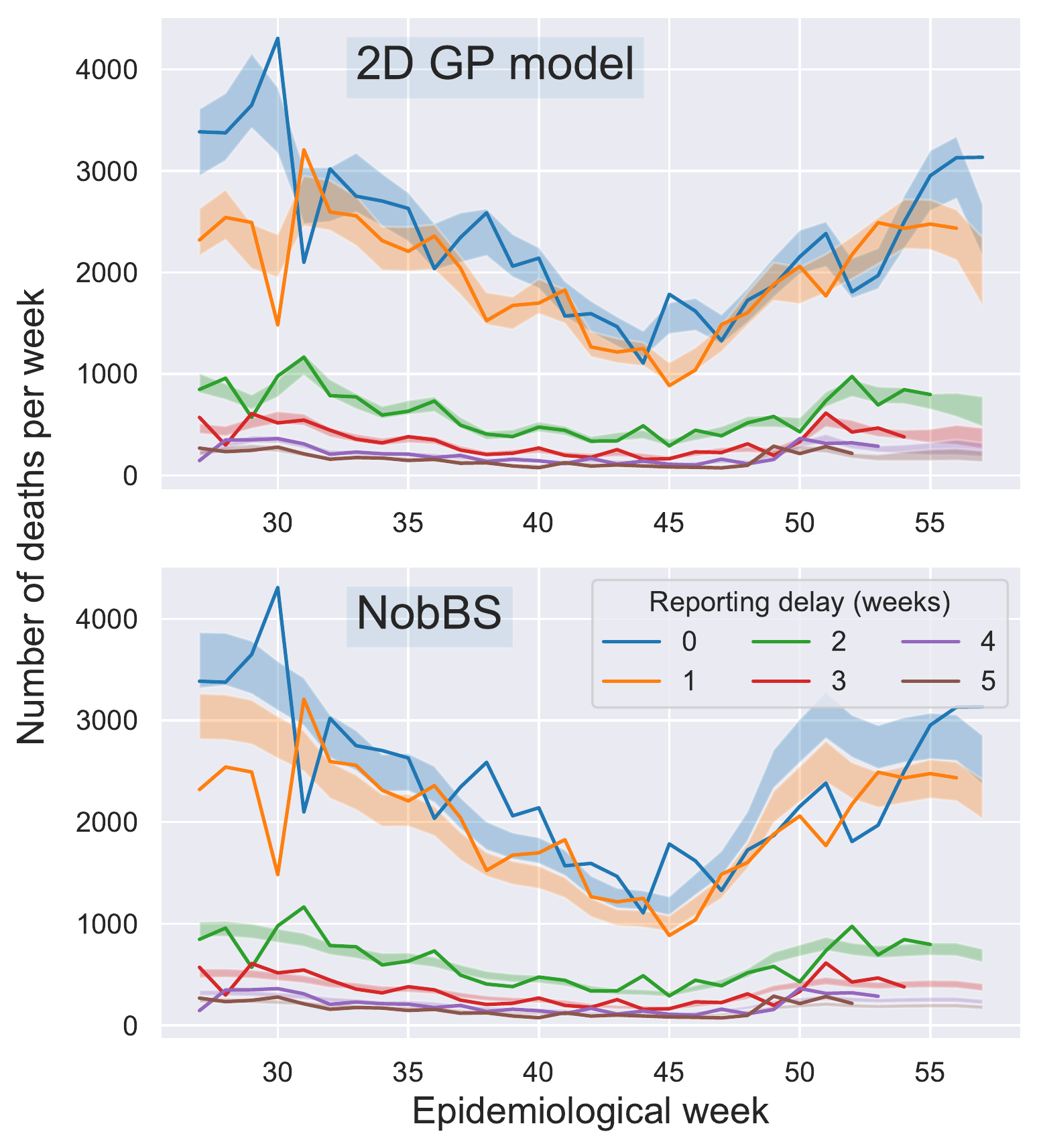}
\par\end{centering}
\caption{Reported and nowcasted numbers of deaths with reporting delay 0 to 5 weeks generated by the 2D additive GP and NobBS models. These plots show the columns of the reporting triangle $n_{t,d}$. The reported data are shown with solid lines, and the 50\% CrI for the nowcasts with the ribbons. 95\% CrI are shown in Figure~\ref{fig: nowcast_brazil_delays_95crI}. \label{fig: nowcast_brazil_delays}}
\end{figure}

Other nowcasting methods, including NobBS, focus primarily on estimating only the "missing" part of the $n_{t,d}$ array and comparing the total numbers $n_t$, that is sums of each row of the array. Here, we aim to obtain a statistical model explaining all elements of the $n_{t,d}$ matrix. The reason for that is twofold: firstly, having a model that describes the whole $n_{t,d}$ surface well increases the reliability of the model, which is vital in any healthcare setting. Secondly, the SIVEP-Gripe database contains hard to identify errors discussed in section~\ref{sec:data}, therefore it is preferable to treat the reported data with additional statistical uncertainty. The fit of the 2D GP and the NobBS models to the $n_{t,d}$ matrix is presented in Figure~\ref{fig: nowcast_brazil_delays} and shows that the GP-based nowcasting method fits the time-delay structure much closer than NobBS.

\subsection{Retrospective testing} \label{sec:testing}

\begin{figure}[!h]
\begin{centering}
\includegraphics[scale=0.5]{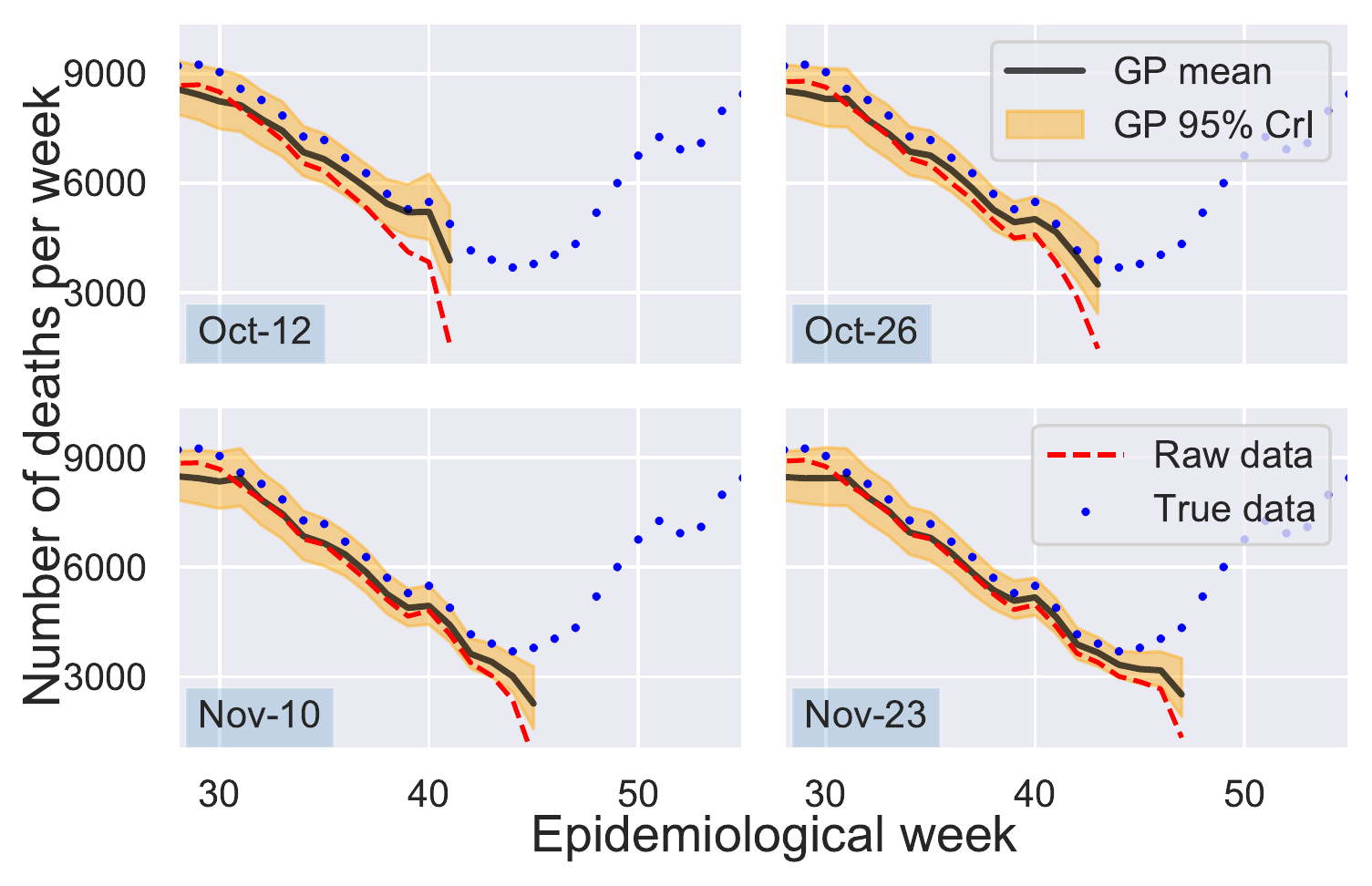}
\par\end{centering}
\caption{Retrospective testing for the whole of Brazil using the 2D additive GP model. \label{fig: brazil_backtest}}
\end{figure}

To evaluate the accuracy of the competing nowcasting models, we fit all of the models to the retrospective data sets available at each week between 05-Oct and 30-Nov-2020, using the numbers of deaths recorded for the whole of Brazil. This way we obtain 63 different sets of nowcasts, which we compare to the numbers of deaths reported by the most recent SIVEP-Gripe data release from 31-May-2021. The start date gives us at least 15 weeks of training data for each of the nowcasts. The end date of 30-Nov is 26 weeks before the most recent release, so the number of deaths reported in the most recent release can be confidently taken as a true value. The comparison is done by calculating the weighted and unweighted rooted mean squared error (RMSE) and the continuous ranked probability score (CRPS) between the "true" values from the most recent release and the nowcasted values. The differences between the ground truth and raw data, as shown in Figure \ref{fig: Rt}A, are used as weights. For each RMSE and CRPS evaluation, we use the mortality data from the 10 weeks leading up to the date of the nowcast. 

\begin{figure}[!ht]
\begin{centering}
\includegraphics[scale=0.25]{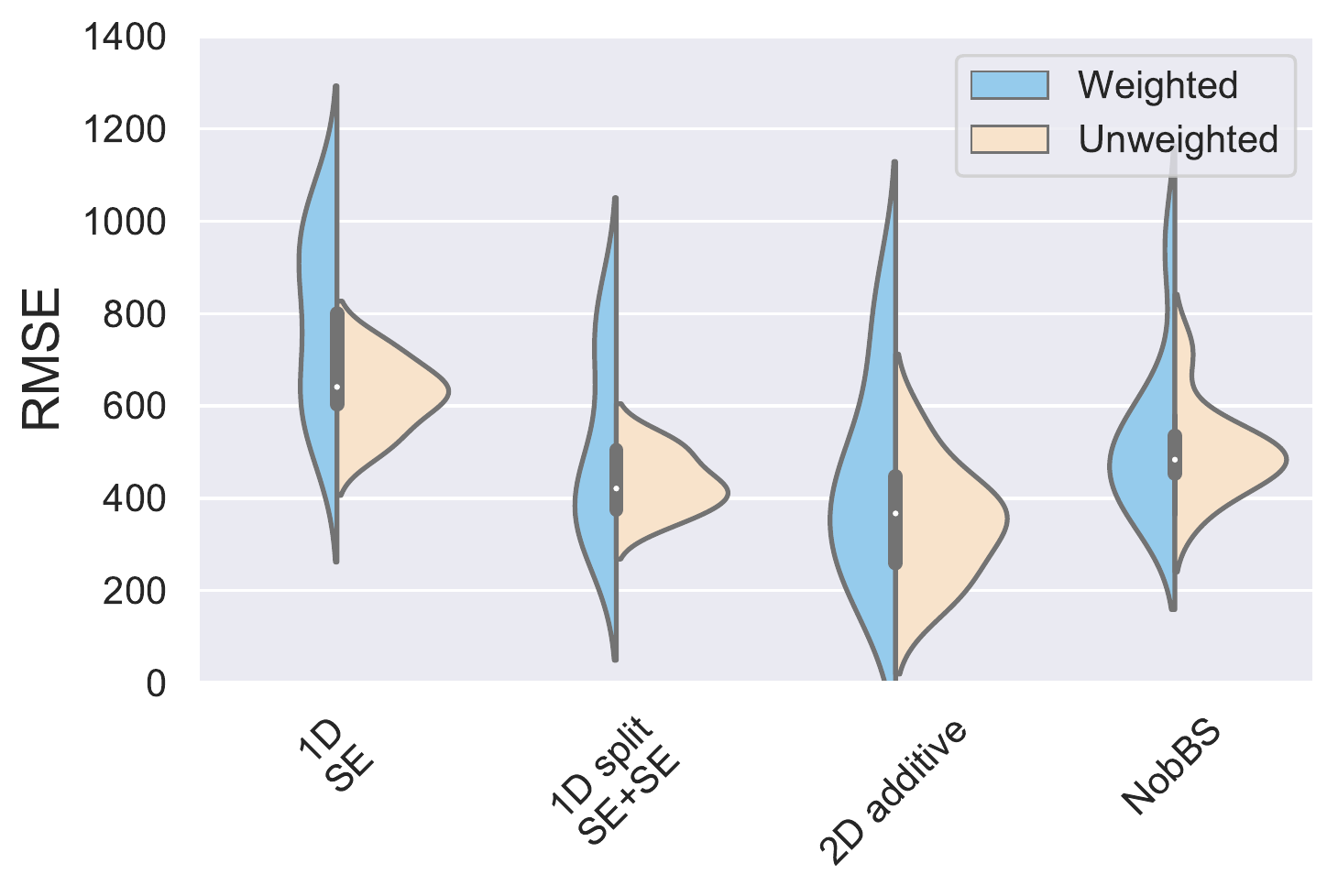}
\includegraphics[scale=0.25]{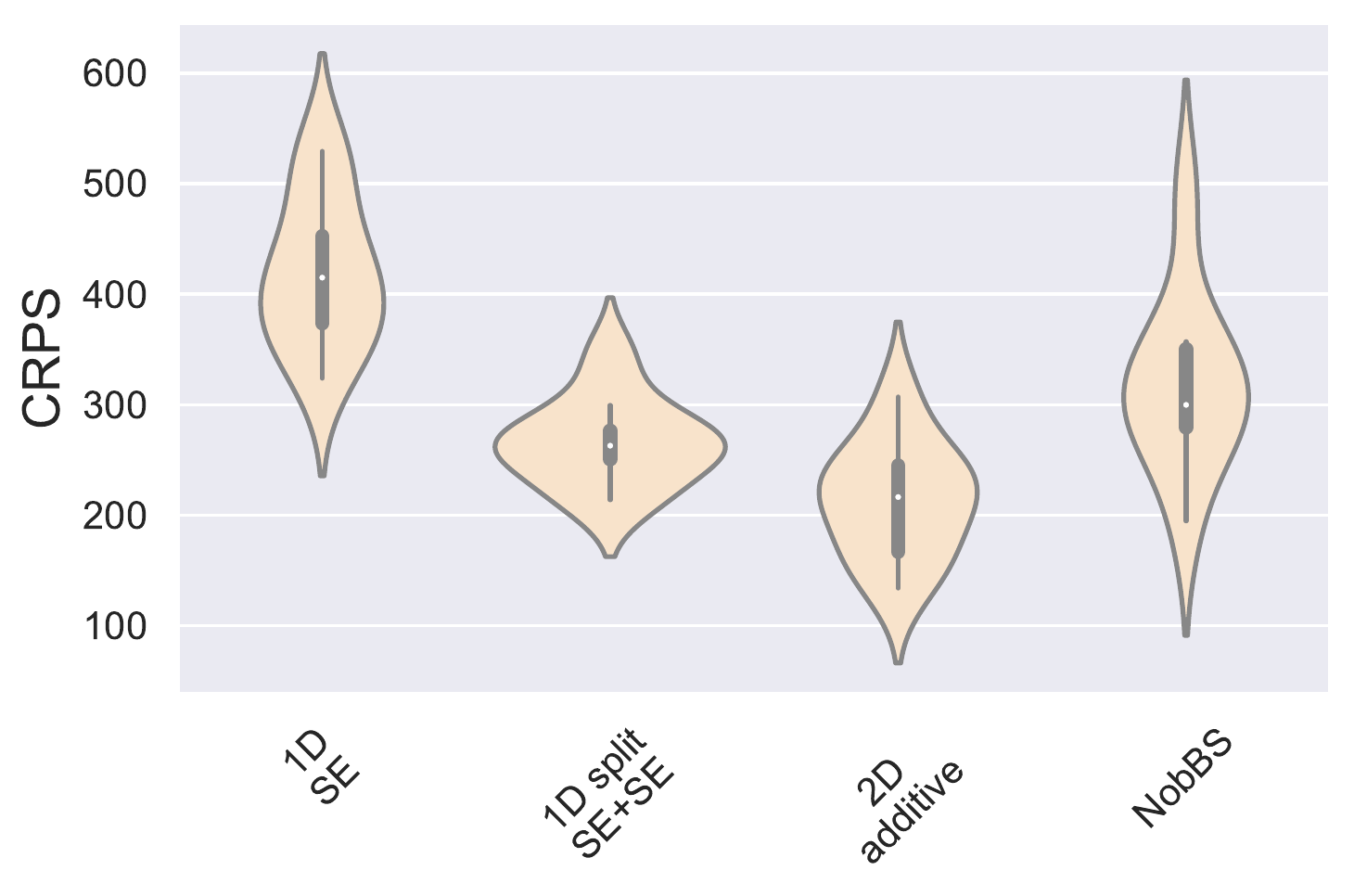}
\includegraphics[scale=0.25]{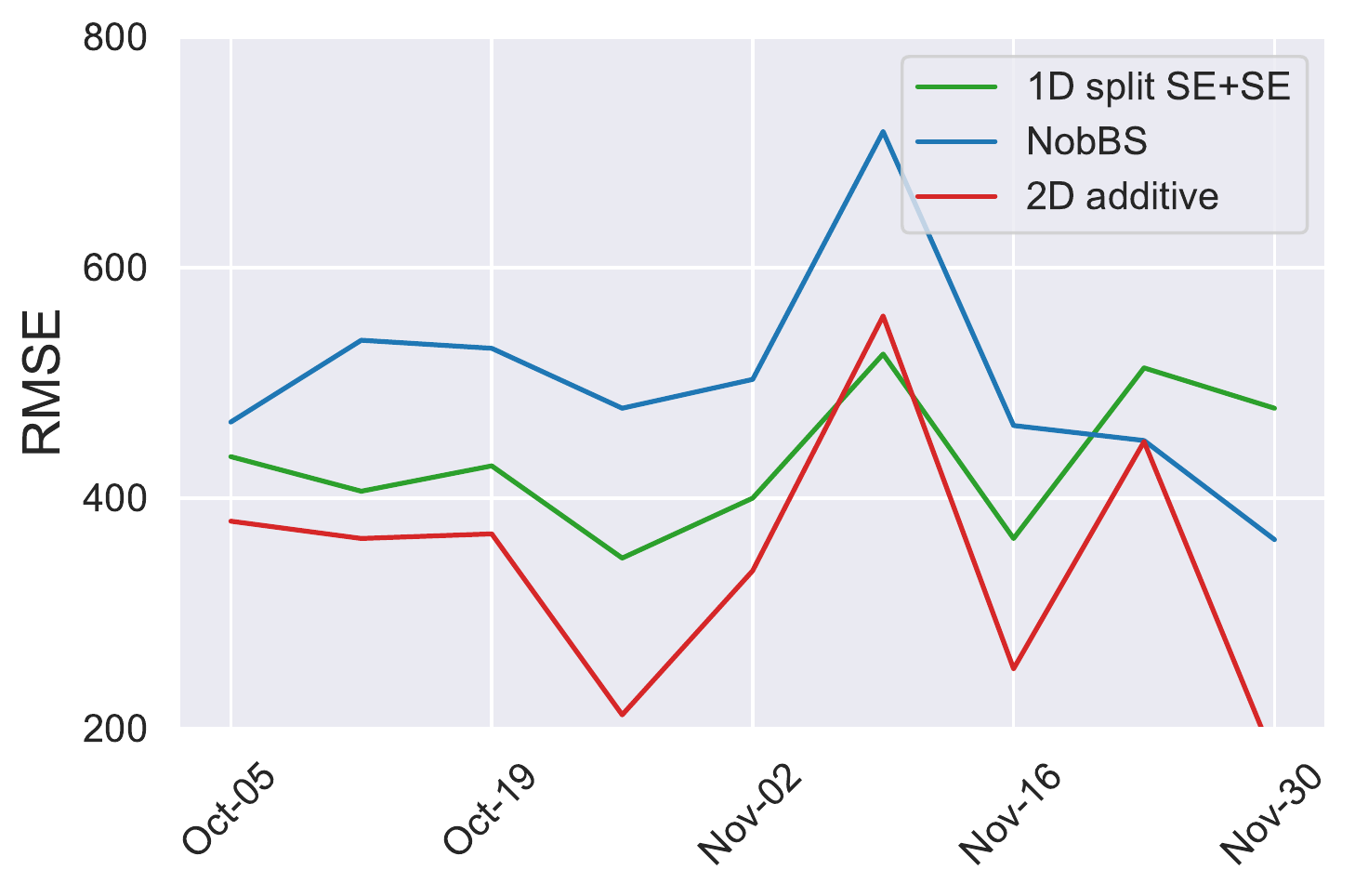}
\includegraphics[scale=0.25]{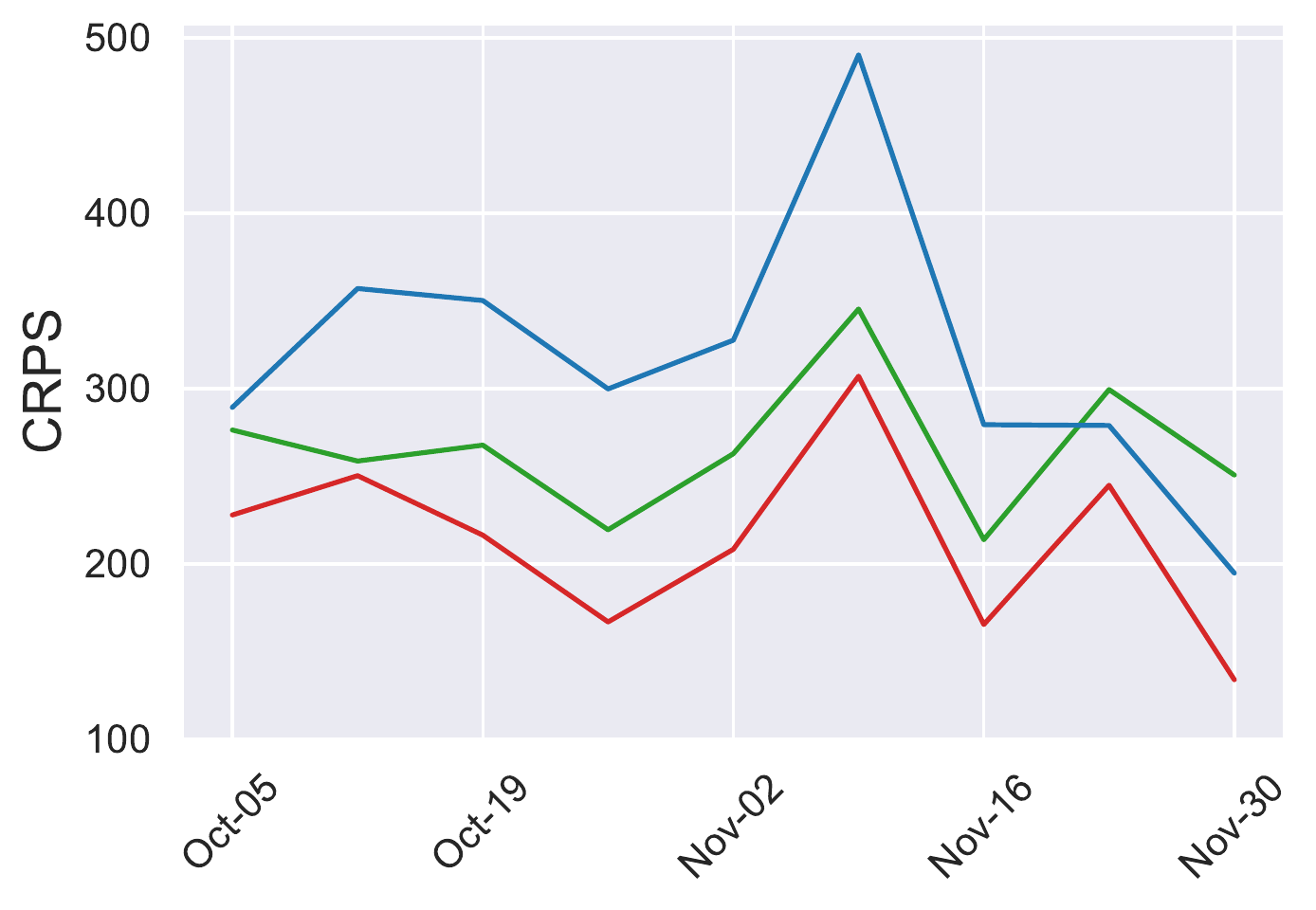}
\par\end{centering}
\caption{RMSE and CRPS evaluated for $n_t$ for tested nowcasting methods over the weeks 5-Oct to 30-Nov-2020. For the weighted RMSE, the weights were calculated as a difference between the ground truth and the raw data available up to the nowcasting date.}\label{fig: RMSE_weighted}
\end{figure}

\begin{figure*}[!h]
\begin{centering}
\includegraphics[scale=0.5]{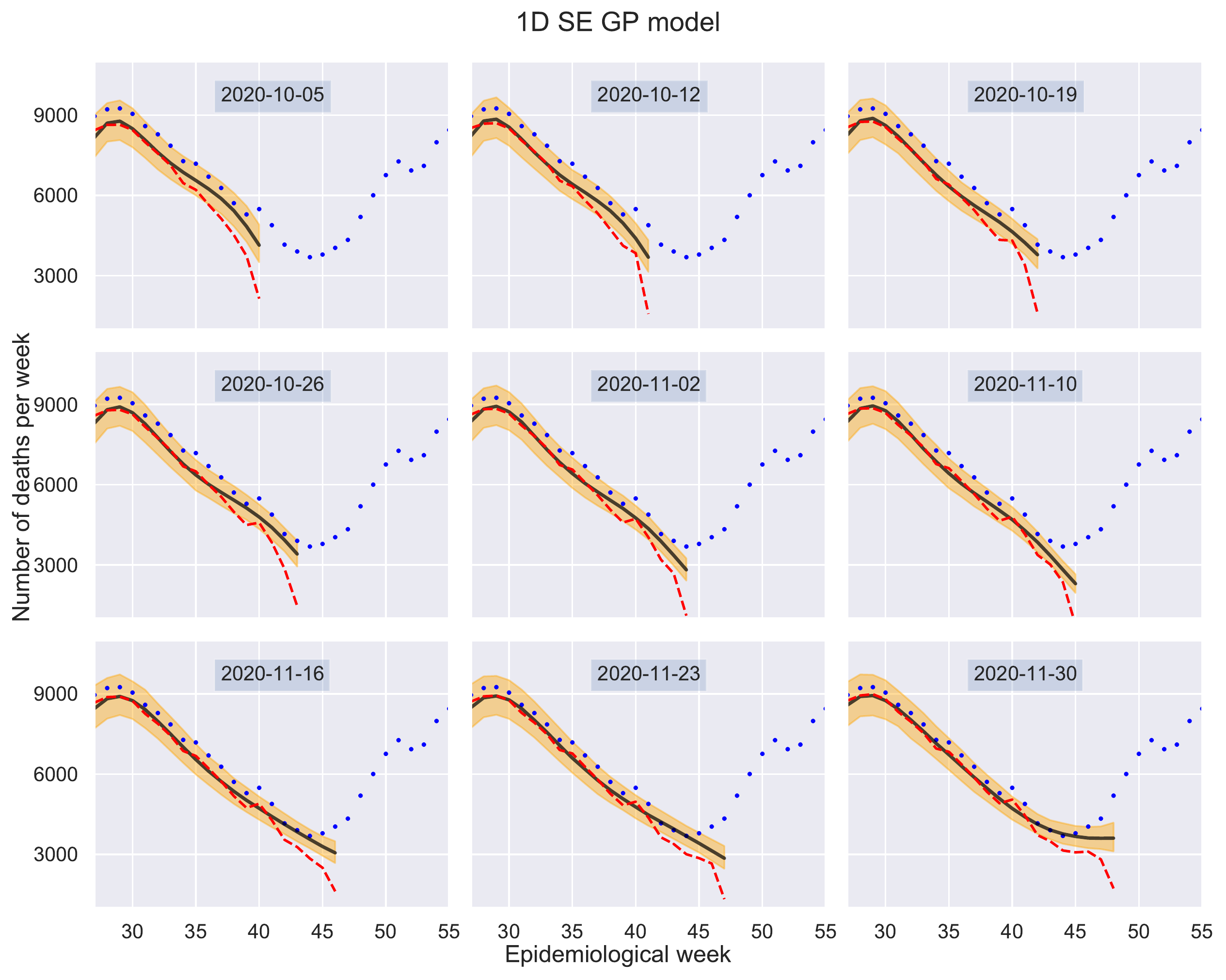}
\par\end{centering}
\caption{Retrospective tests for the 1D SE GP nowcasting mode. Deaths reported in the 31-May-2021 release are shown with blue dots, data available at the time of nowcasting with red dashed line, nowcasted mean values with black solid line and 95\% CrI with orange ribbon. \label{fig: all_dates_1}}
\end{figure*}

\begin{figure*}[!h]
\begin{centering}
\includegraphics[scale=0.5]{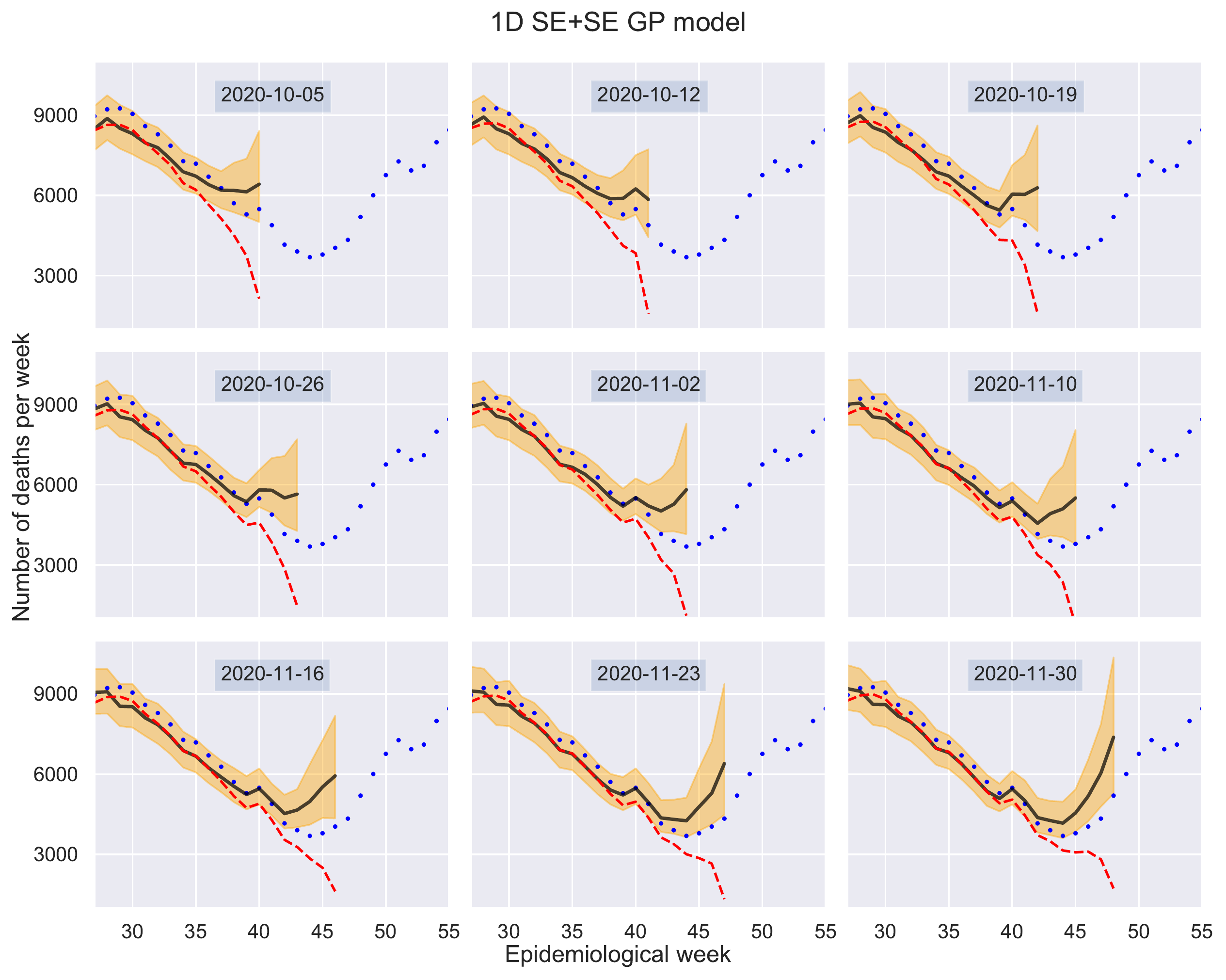}
\par\end{centering}
\caption{Retrospective tests for the 1D SE+SE GP model. \label{fig: all_dates_2}}
\end{figure*}

\begin{figure*}[!h]
\begin{centering}
\includegraphics[scale=0.5]{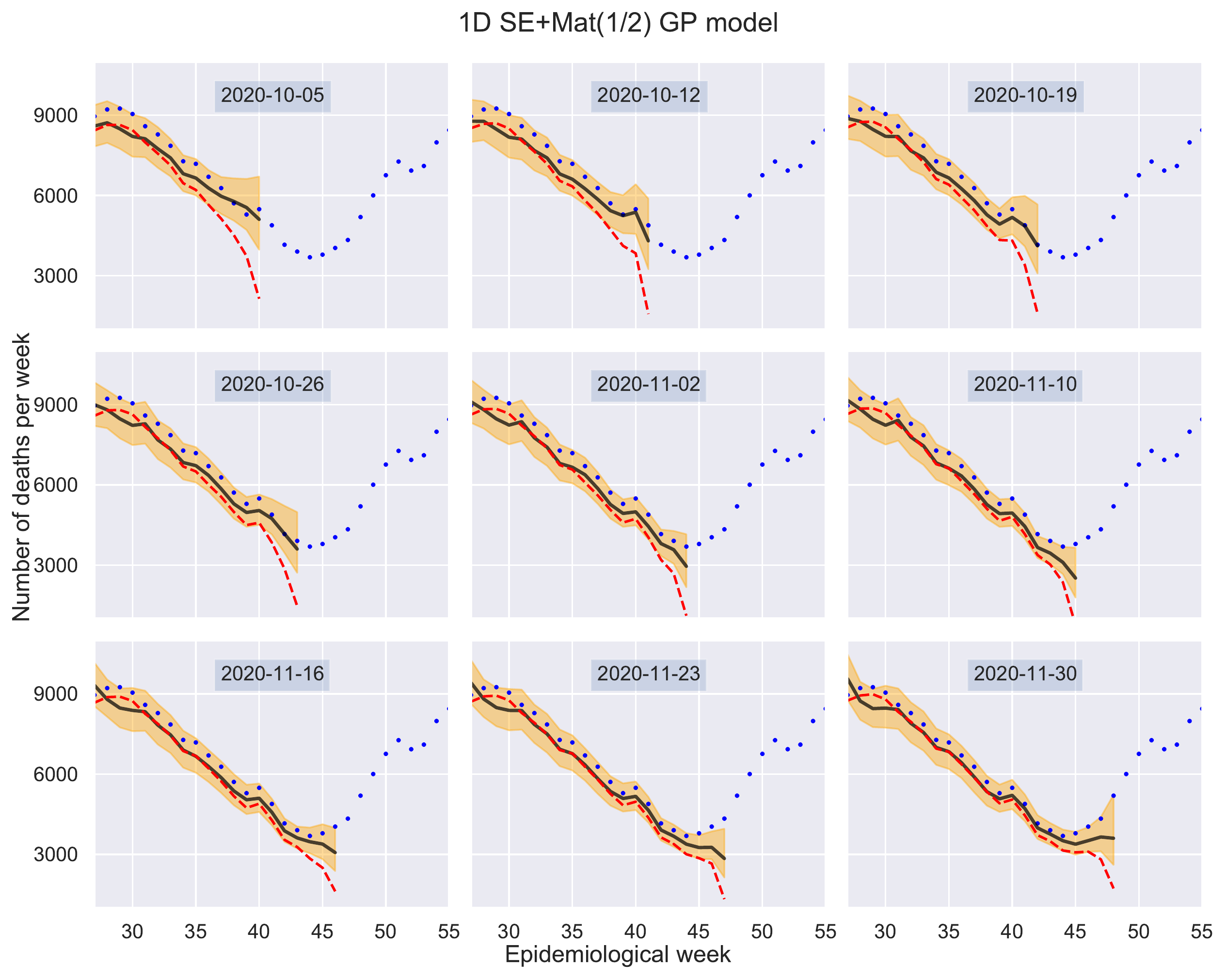}
\par\end{centering}
\caption{Retrospective tests for the 1D SE+Mat(1/2) GP model. \label{fig: all_dates_3}}
\end{figure*}

\begin{figure*}[!h]
\begin{centering}
\includegraphics[scale=0.5]{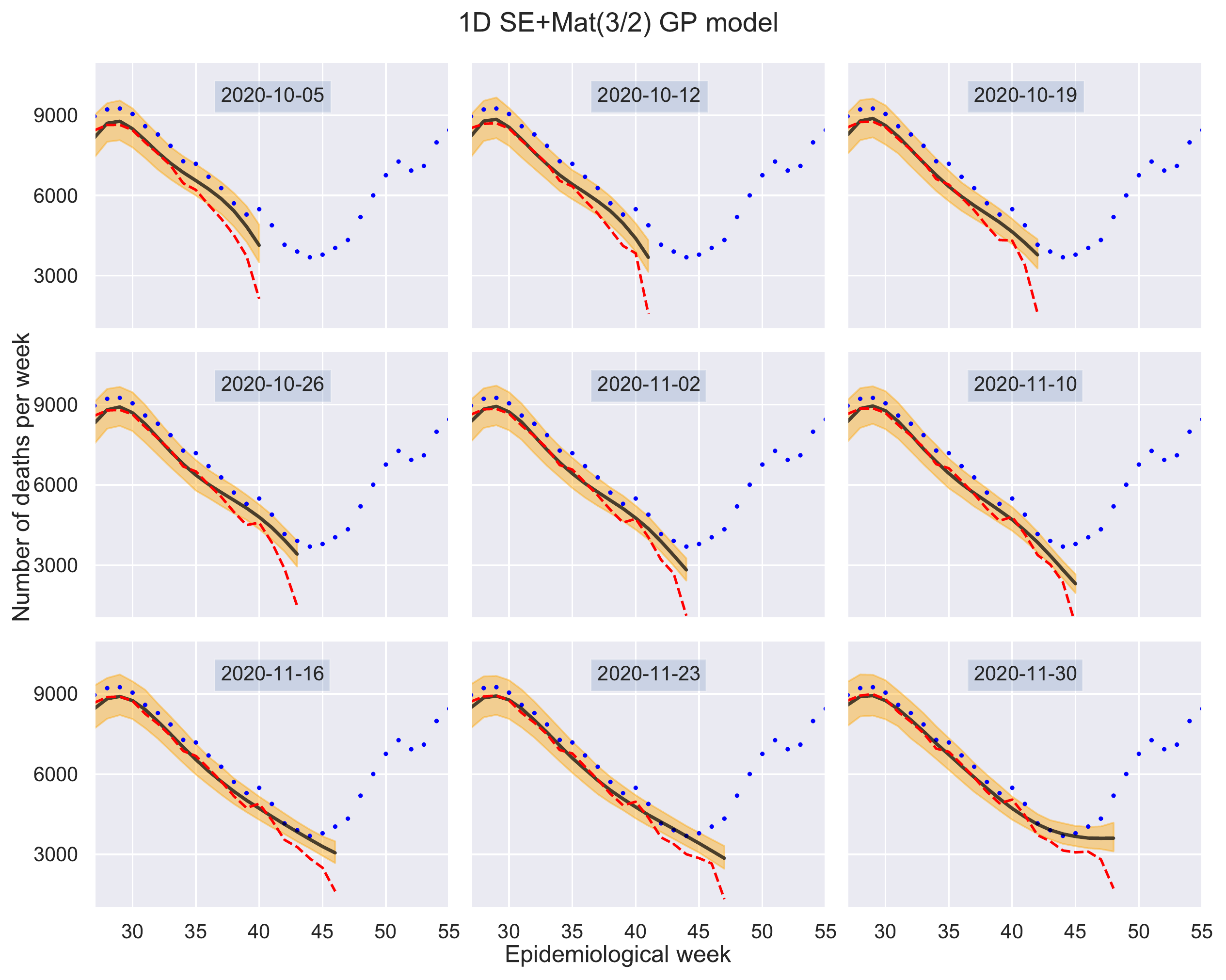}
\par\end{centering}
\caption{Retrospective tests for the 1D SE+Mat(3/2) GP model. \label{fig: all_dates_4}}
\end{figure*}

\begin{figure*}[!h]
\begin{centering}
\includegraphics[scale=0.5]{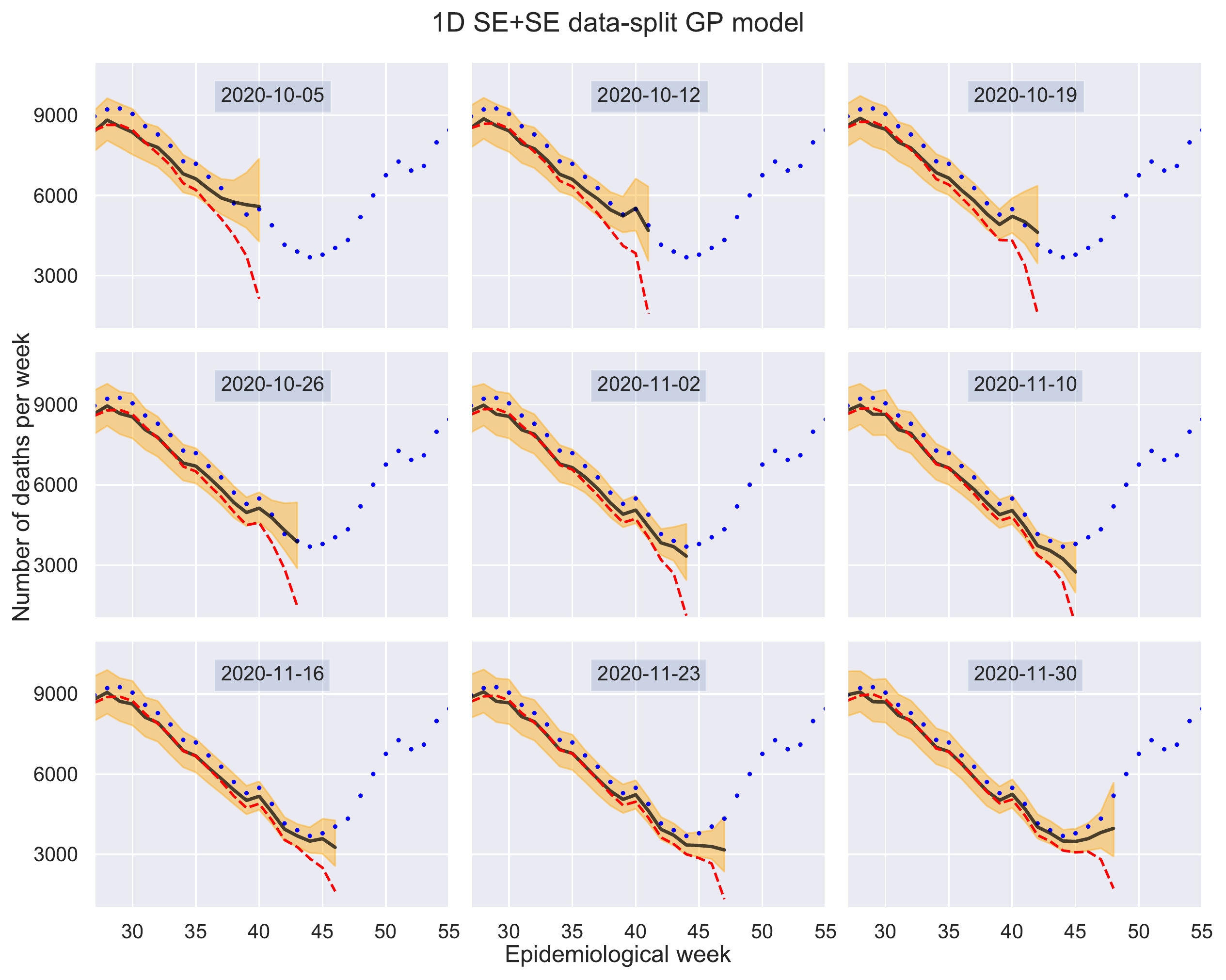}
\par\end{centering}
\caption{Retrospective tests for the 1D SE+SE data-split GP model. \label{fig: all_dates_7}}
\end{figure*}

Out of all tested models, GP models with 1D kernels with 2 components (SE+SE and SE+Mat) performed worse than the benchmark. The predictive accuracy of the other models was comparable to that of the benchmark, as shown in Figure~\ref{fig: RMSE_weighted}, while also simultaneously giving an appropriate statistical description of the data (Figure~\ref{fig: nowcast_brazil_delays}). These results provide empirical evidence that our proposed method, under correct specification, gives a complete and accurate approach for describing and nowcasting COVID-19 death data. Model fits for the 2D additive GP model, including the 95\% credible intervals (CrI) are shown in Figure~\ref{fig: brazil_backtest} and the fits for the remaining models are shown in Figures \ref{fig: all_dates_1} - \ref{fig: all_dates_10}.

\begin{figure}[!htb]
\begin{centering}
\includegraphics[scale=0.45]{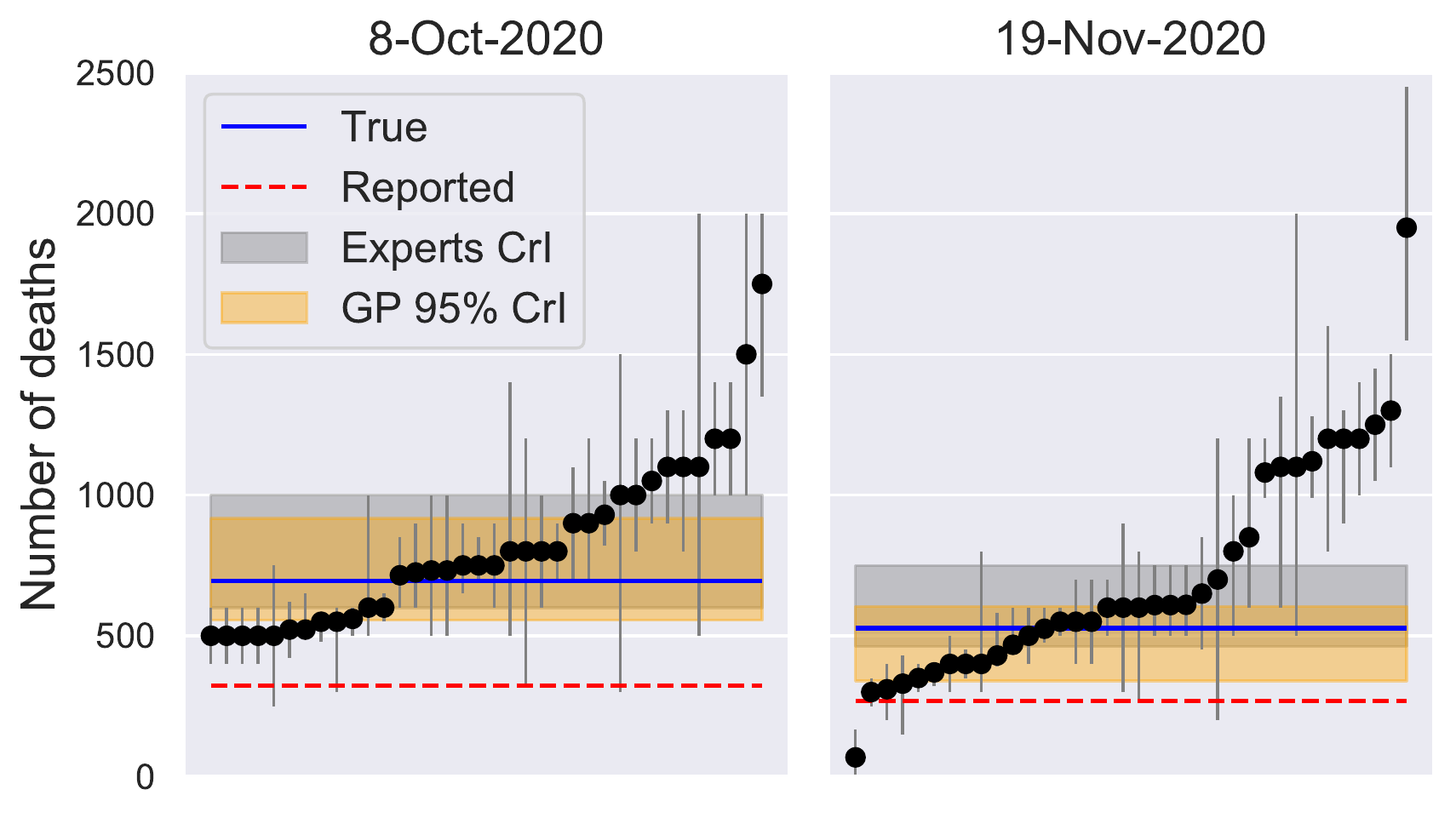}
\par\end{centering}
\caption{Human experts estimates of the true number of deaths are shown with the black points and errorbars. \label{fig: experts}}
\end{figure}

\begin{figure*}[!h]
\begin{centering}
\includegraphics[scale=0.5]{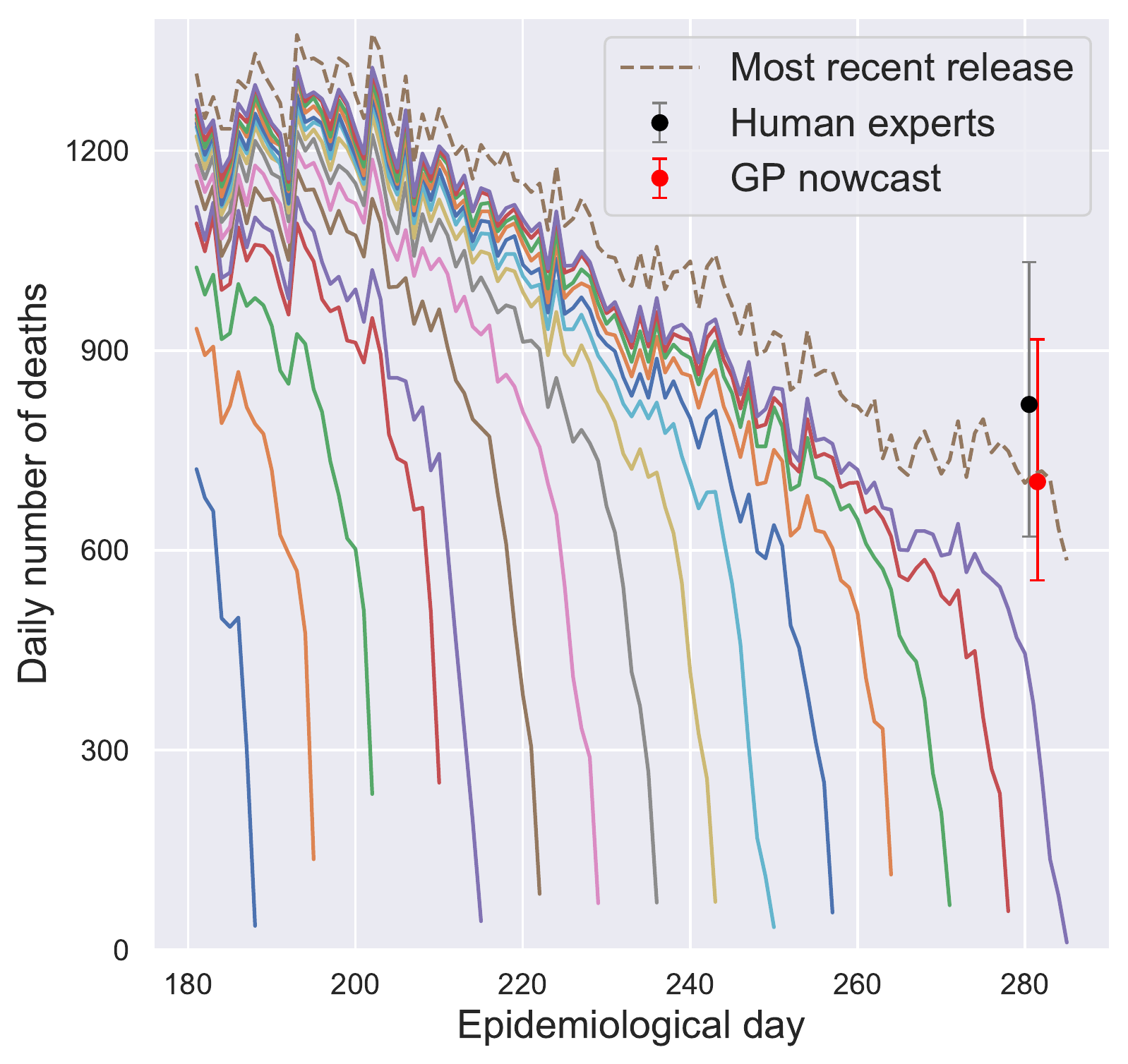}
\includegraphics[scale=0.5]{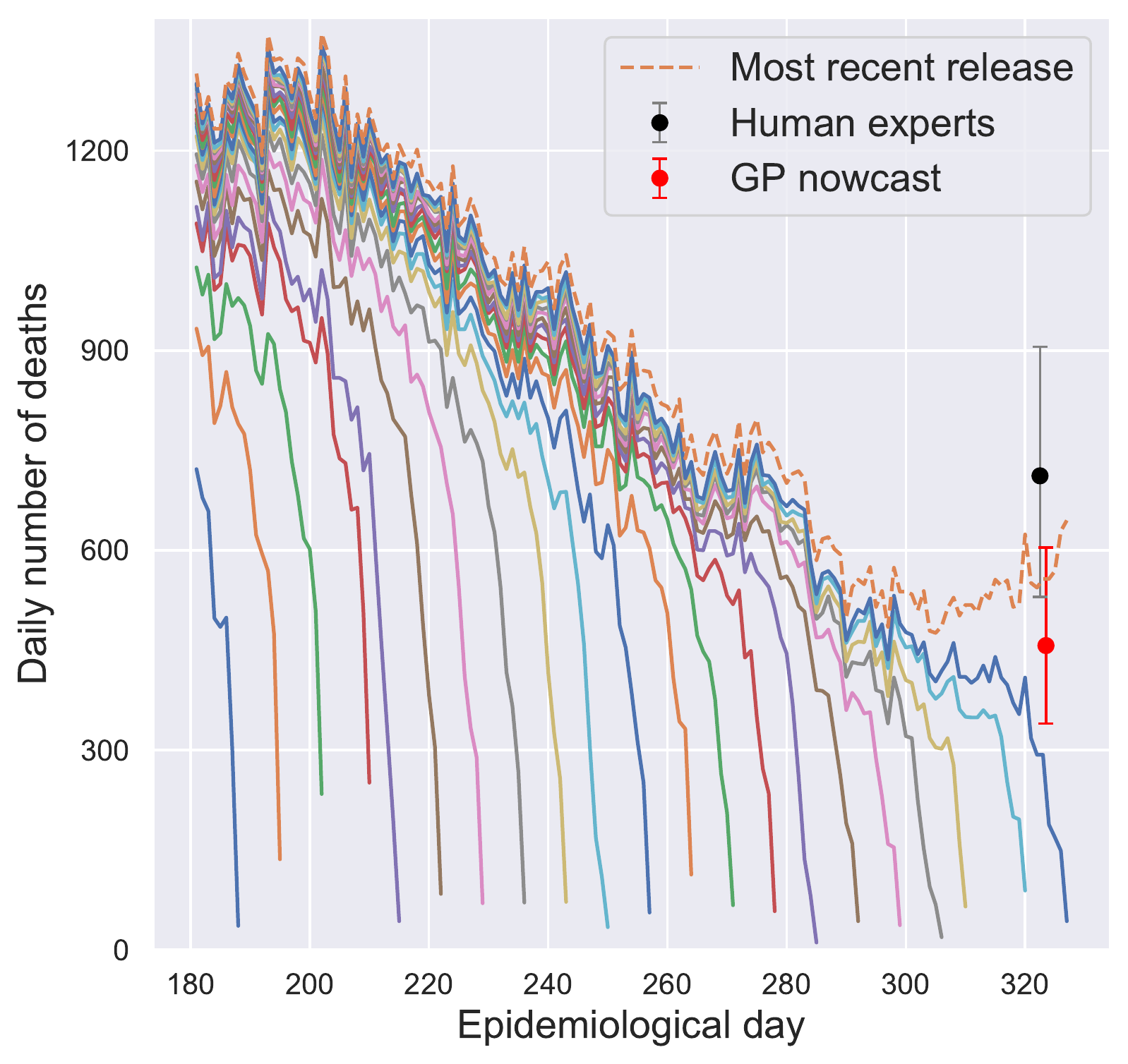}
\par\end{centering}
\caption{Human experts and 1D SE+SE data-split GP model estimates of a true number of deaths on 8-Oct-2020 (left) and 19-Nov-2020 (right) plotted together with the reported data. \label{fig: SMexperts}}
\end{figure*}

\begin{figure}[!h]
\begin{centering}
\includegraphics[scale=0.5]{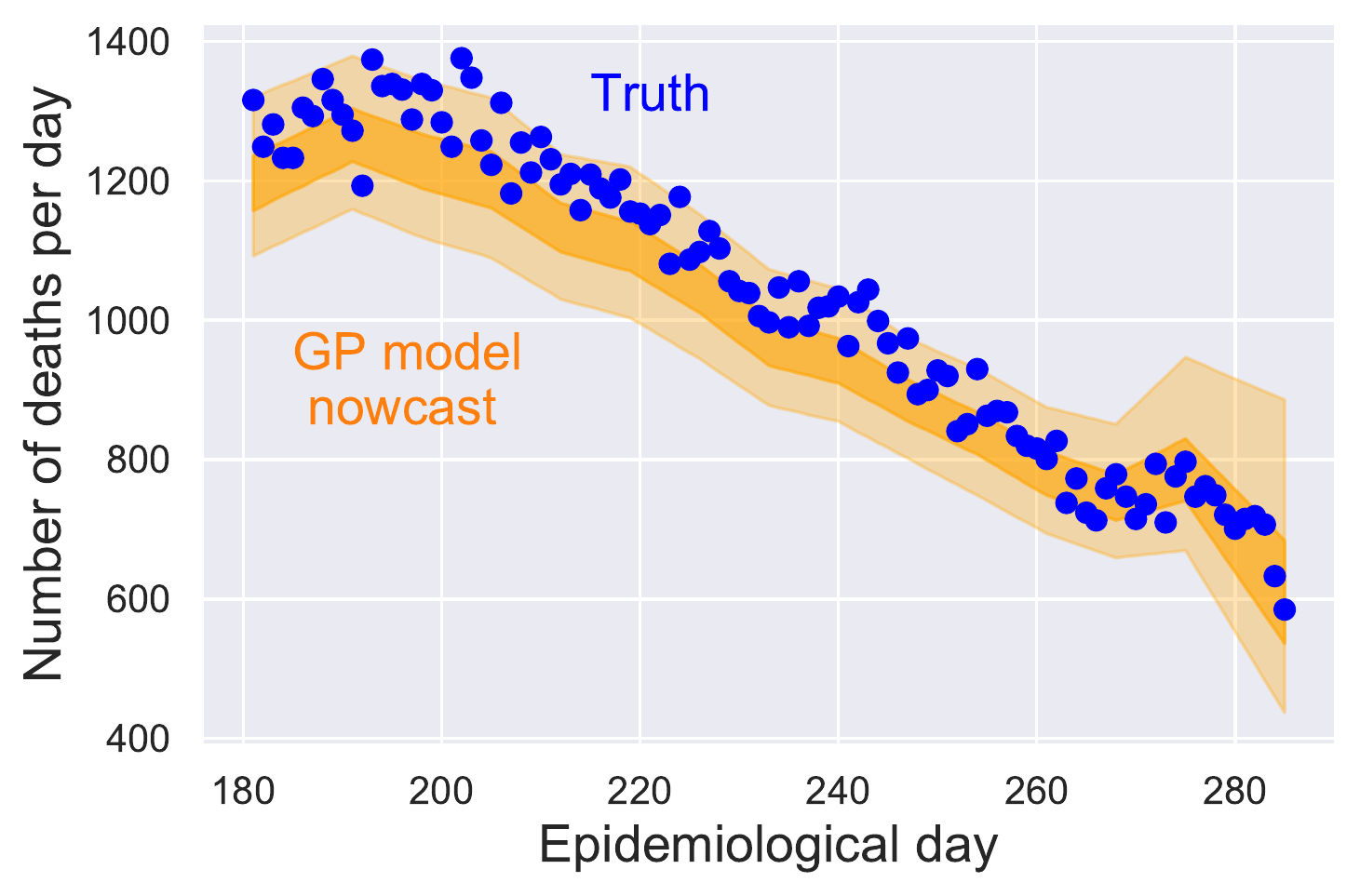}
\par\end{centering}
\caption{Example of the interpolation of numbers of deaths per day based on the weekly nowcasted values. The 50\% and 95\% CrI for the nowcast are shown with the ribbon. \label{fig: daily_interp}}
\end{figure}

In addition to the forecasting metrics, as a novelty, we also evaluate how the GP nowcasting model performs when compared to human experts' predictions. We asked a group of infectious disease epidemiology experts to provide a series of nowcasts when presented with time series data up to 12-Oct and 23-Nov-2020. They were asked for their estimates of the true numbers of deaths due to COVID-19 in Brazil on the 08-Oct and 19-Nov-2020 (Figure~\ref{fig: SMexperts}). The dates were specifically chosen to represent different scenarios. In the first one, 08-Oct-2020, both the raw data and the updated numbers of daily deaths were declining. Whereas in the second date, 19-Nov-2020, the raw data were declining while the updated release revealed that the true numbers of daily deaths were actually increasing. For this experiment, 36 anonymous experts provided their point estimates and confidence intervals, which are presented in Figure~\ref{fig: experts}. To extract daily deaths from the model's weekly estimates, we performed a simple interpolation, through setting the nowcasted number of deaths per week divided by 7 to the middle day of the given week, and interpolating the remaining values using splines (example shown in Figure~\ref{fig: daily_interp}).

Notably in both cases the median value guessed by the human experts was not far from the "true" value (difference of 55 and 73 deaths/day respectively for the human median estimate and 13 and 72 for the nowcasting model mean), however only 36\% and 50\% of all answers included the true value within the provided credible intervals, respectively. The confidence intervals given by the human experts were comparable with the 95\% CrI of the model, with the model confidence narrower by 19 deaths/day for the first date and 27 deaths/day for the second date.

\subsection{Sensitivity analysis}
\begin{figure}[!h]
\begin{centering}
\includegraphics[scale=0.6]{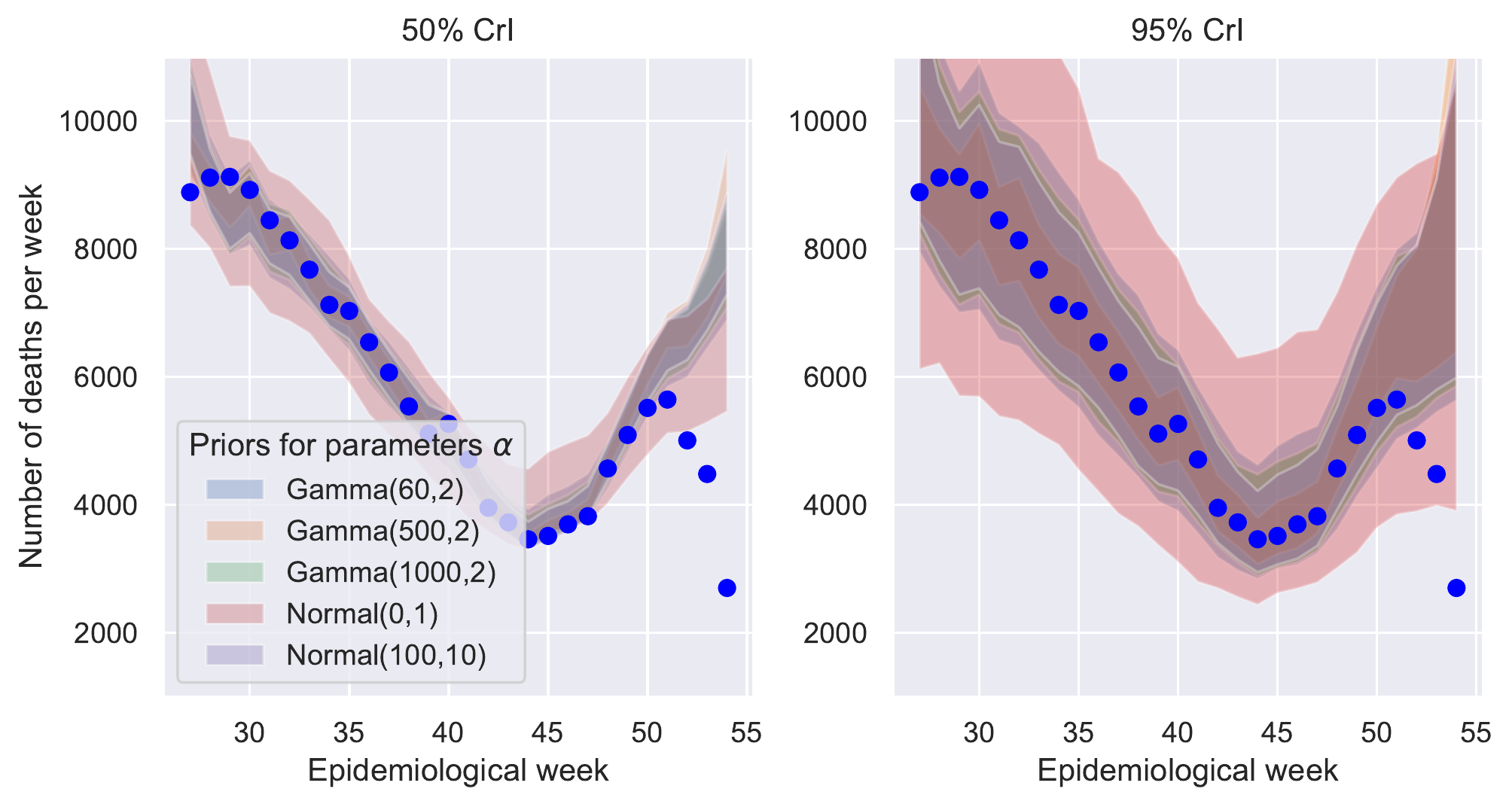}
\par\end{centering}
\caption{Model fits with different $r$ prior density. \label{fig: sensitivity_r}}
\end{figure}

\begin{figure}[!h]
\begin{centering}
\includegraphics[scale=0.4]{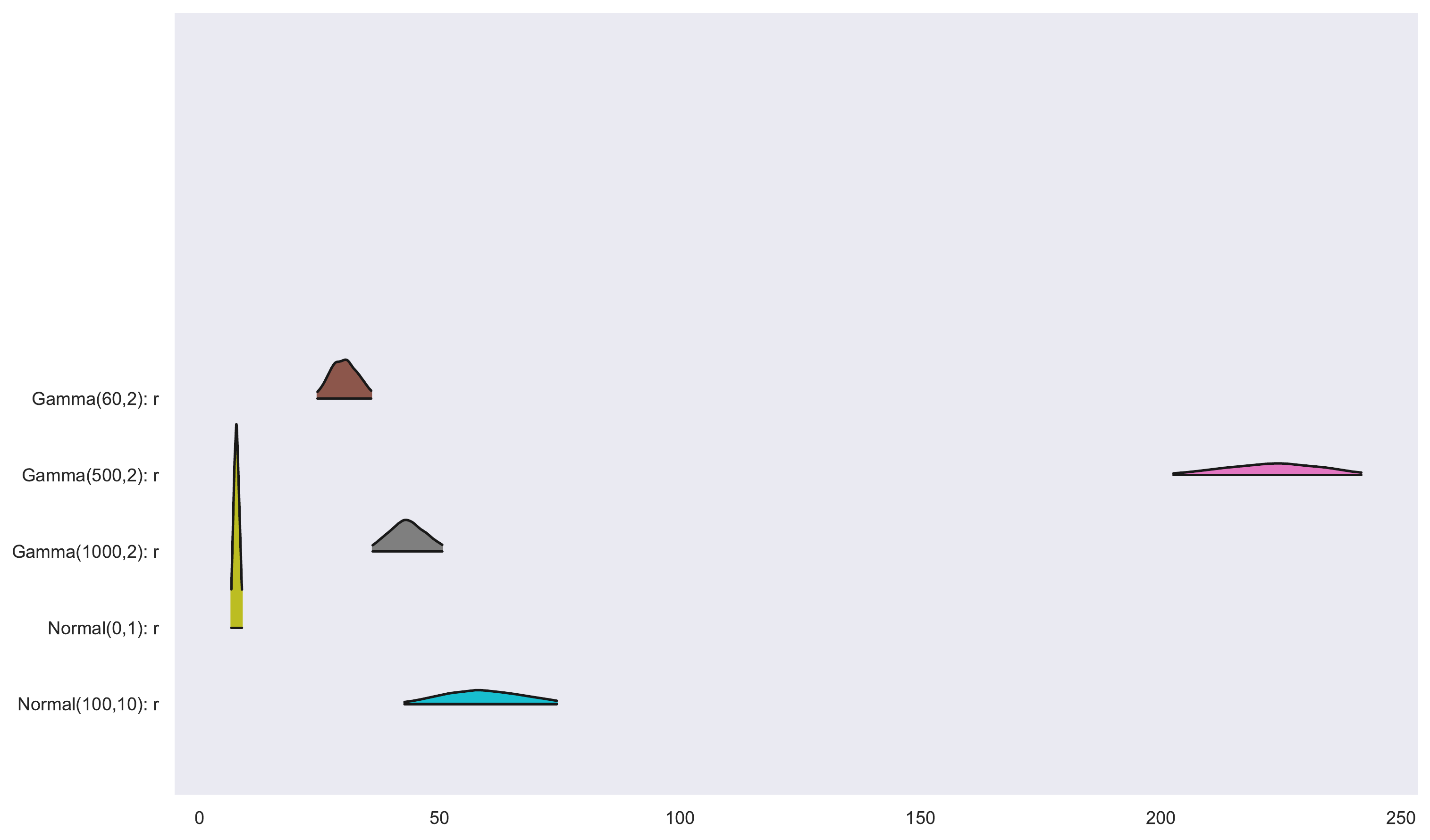}
\par\end{centering}
\caption{Model fits with different $r$ prior density.  \label{fig: sensitivity_r_ridgeplots}}
\end{figure}

\begin{figure*}[!h]
\begin{centering}
\includegraphics[scale=0.6]{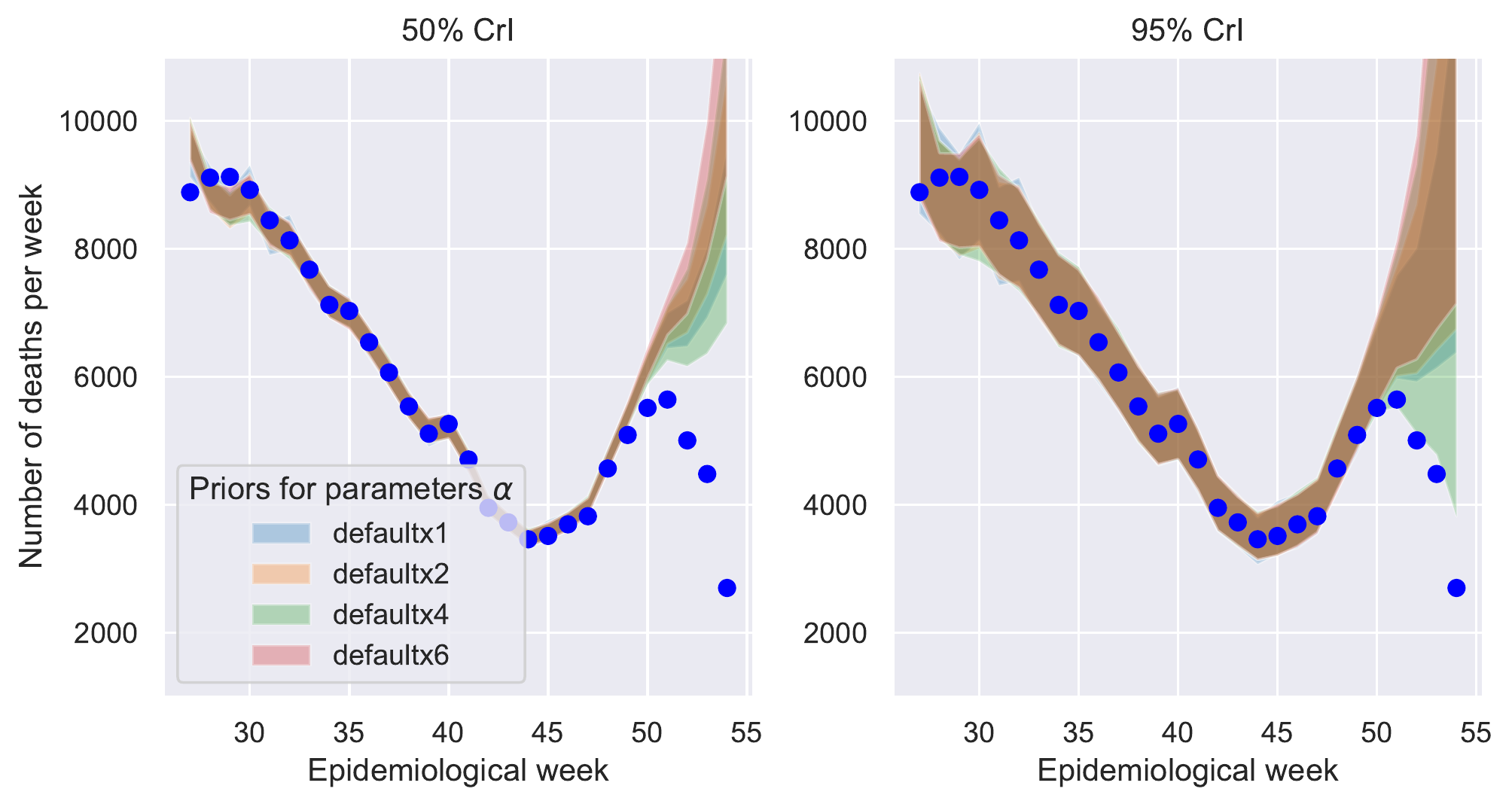}
\par\end{centering}
\caption{Model fits with different $\alpha_{\text{long},1}$, $\alpha_{\text{long},2}$, $\alpha_{\text{short},1}$ and $\alpha_{\text{short},2}$ prior density variance. \label{fig: sensitivity_alpha_var}}
\end{figure*}

\begin{figure*}[!h]
\begin{centering}
\includegraphics[scale=0.6]{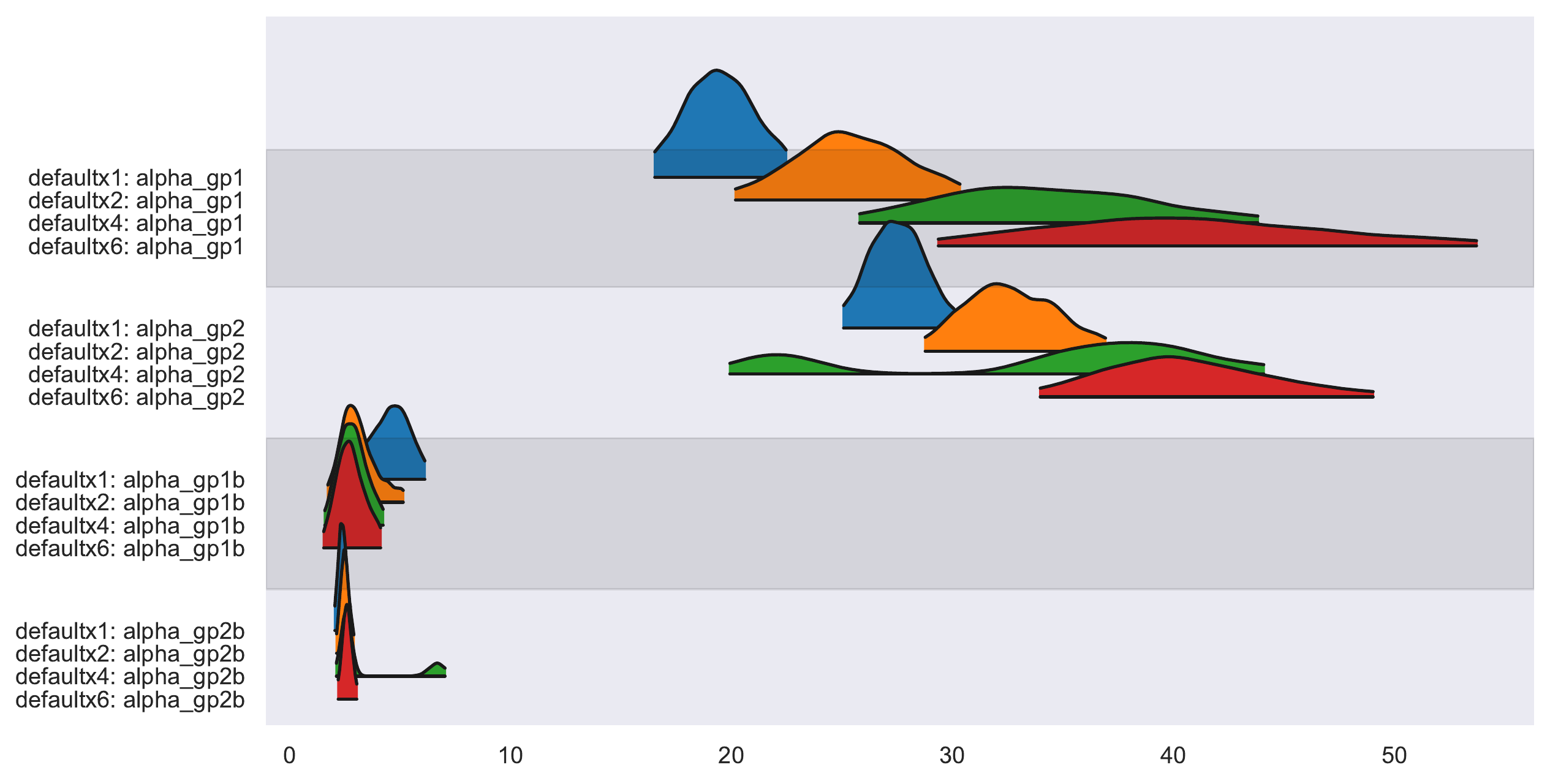}
\par\end{centering}
\caption{Model fits with different $\alpha_{\text{long},1}$, $\alpha_{\text{long},2}$, $\alpha_{\text{short},1}$ and $\alpha_{\text{short},2}$ prior density variance. \label{fig: sensitivity_alpha_var_ridgeplots}}
\end{figure*}

\begin{figure*}[!h]
\begin{centering}
\includegraphics[scale=0.6]{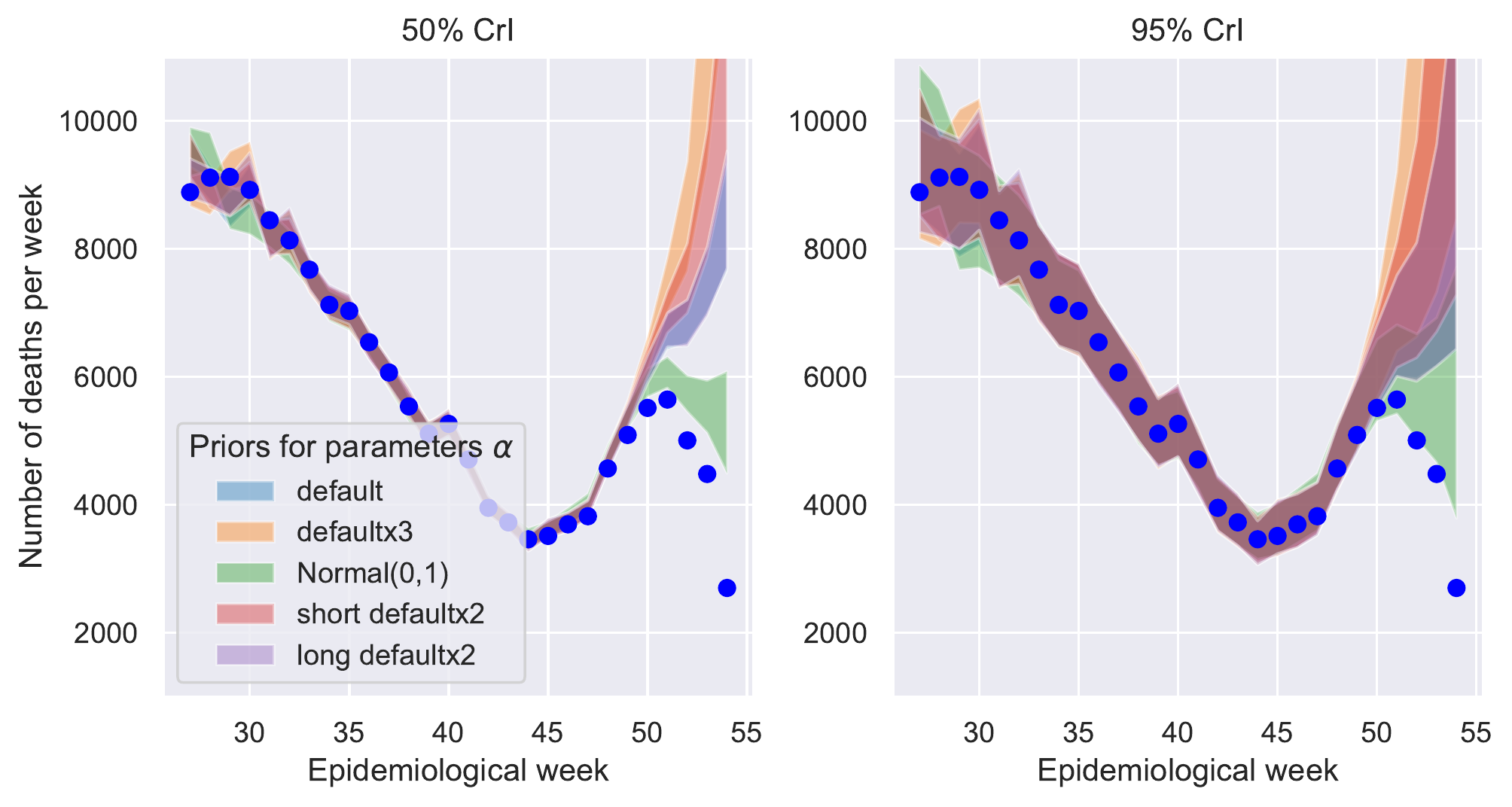}
\par\end{centering}
\caption{Model fits with different $\alpha_{\text{long},1}$, $\alpha_{\text{long},2}$, $\alpha_{\text{short},1}$ and $\alpha_{\text{short},2}$ prior density. Default means using the default priors described in Table~\ref{tab: priors}, for default x 3 prior we increased the mean in the default priors 3-fold, Normal(0,1) means a standard prior was set to all $\alpha$-s, and long- and short- default x 2 means increased mean in the default prior long- and short-part respectively.\label{fig: sensitivity_alpha}}
\end{figure*}

\begin{figure*}[!h]
\begin{centering}
\includegraphics[scale=0.6]{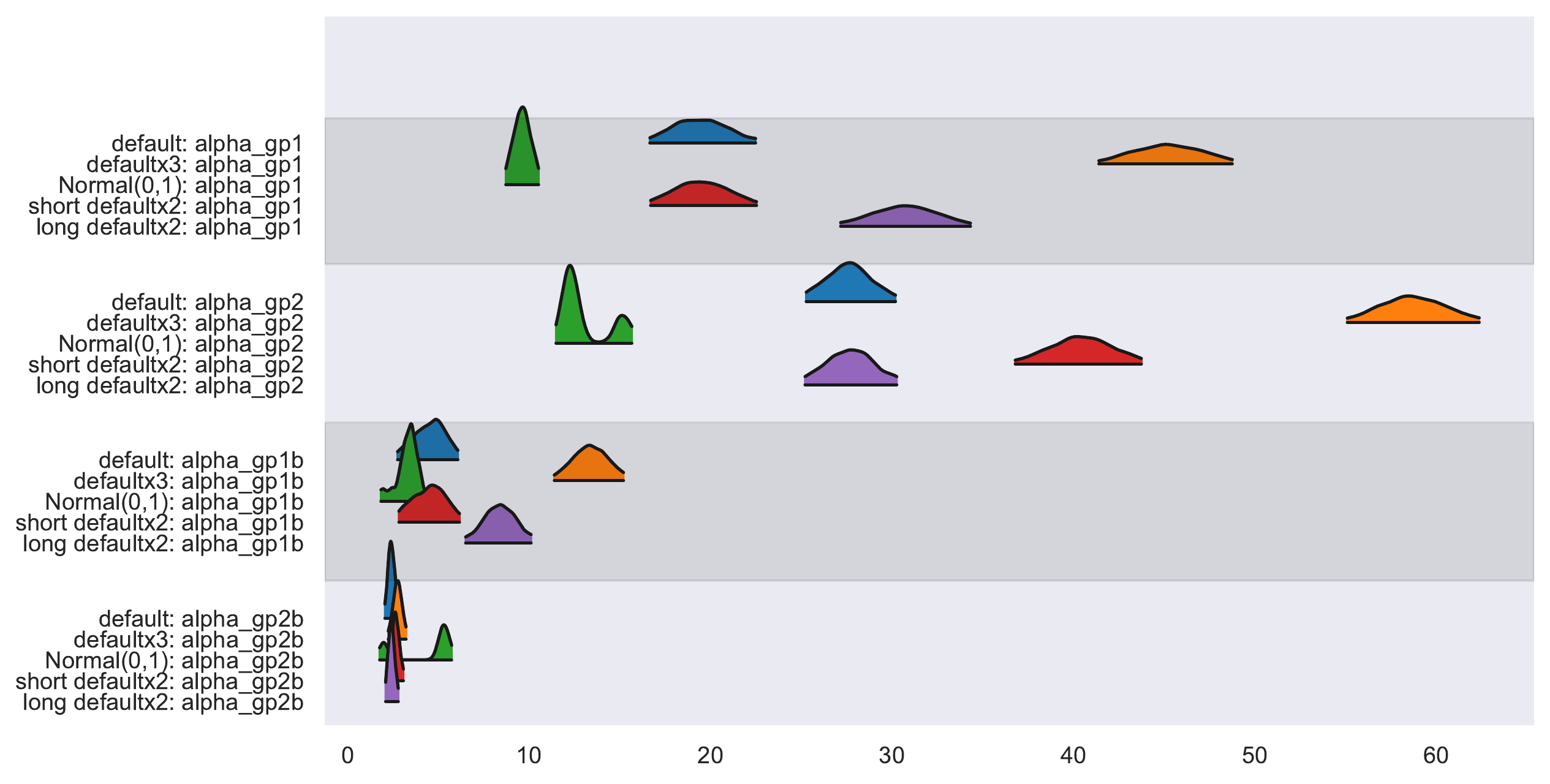}
\par\end{centering}
\caption{Model fits with different $\alpha_{\text{long},1}$, $\alpha_{\text{long},2}$, $\alpha_{\text{short},1}$ and $\alpha_{\text{short},2}$ prior density. Default means using the default priors described in Table~\ref{tab: priors}, for default x 3 prior we increased the mean in the default priors 3-fold, Normal(0,1) means a standard prior was set to all $\alpha$-s, and long- and short- default x 2 means increased mean in the default prior long- and short-part respectively.  \label{fig: sensitivity_alpha_ridgeplots}}
\end{figure*}

We performed basic sensitivity analysis for the 1D SE+SE data-split GP model. We first varied the prior for the overdispersion parameter $r$. This parameter is often unidentifiable by the models and has to be chosen based on the data \citep{mcgough2020nowcasting}. This is confirmed by our sensitivity analysis, where changing the prior for $r$ changed the width of the confidence interval, but did not impact the mean predictions, as shown in Figures~\ref{fig: sensitivity_r} and \ref{fig: sensitivity_r_ridgeplots}.
Changing the priors for the scale parameter $\alpha$ to less informative, that is increasing the variance of the priors does not significantly affect the mean predictions (Figures~\ref{fig: sensitivity_alpha_var}, \ref{fig: sensitivity_alpha_var_ridgeplots}). Changing the mean of the priors does however have an impact on the predictions and leads to bi-modality of the posterior distribution if the model is misspecified, e.g. when $\text{N}(0,1)$ priors are used (Figures~\ref{fig: sensitivity_alpha}, \ref{fig: sensitivity_alpha_ridgeplots}).

\section{Discussion}\label{sec:discussion}
\begin{figure}[!h]
\begin{centering}
\includegraphics[scale=0.3]{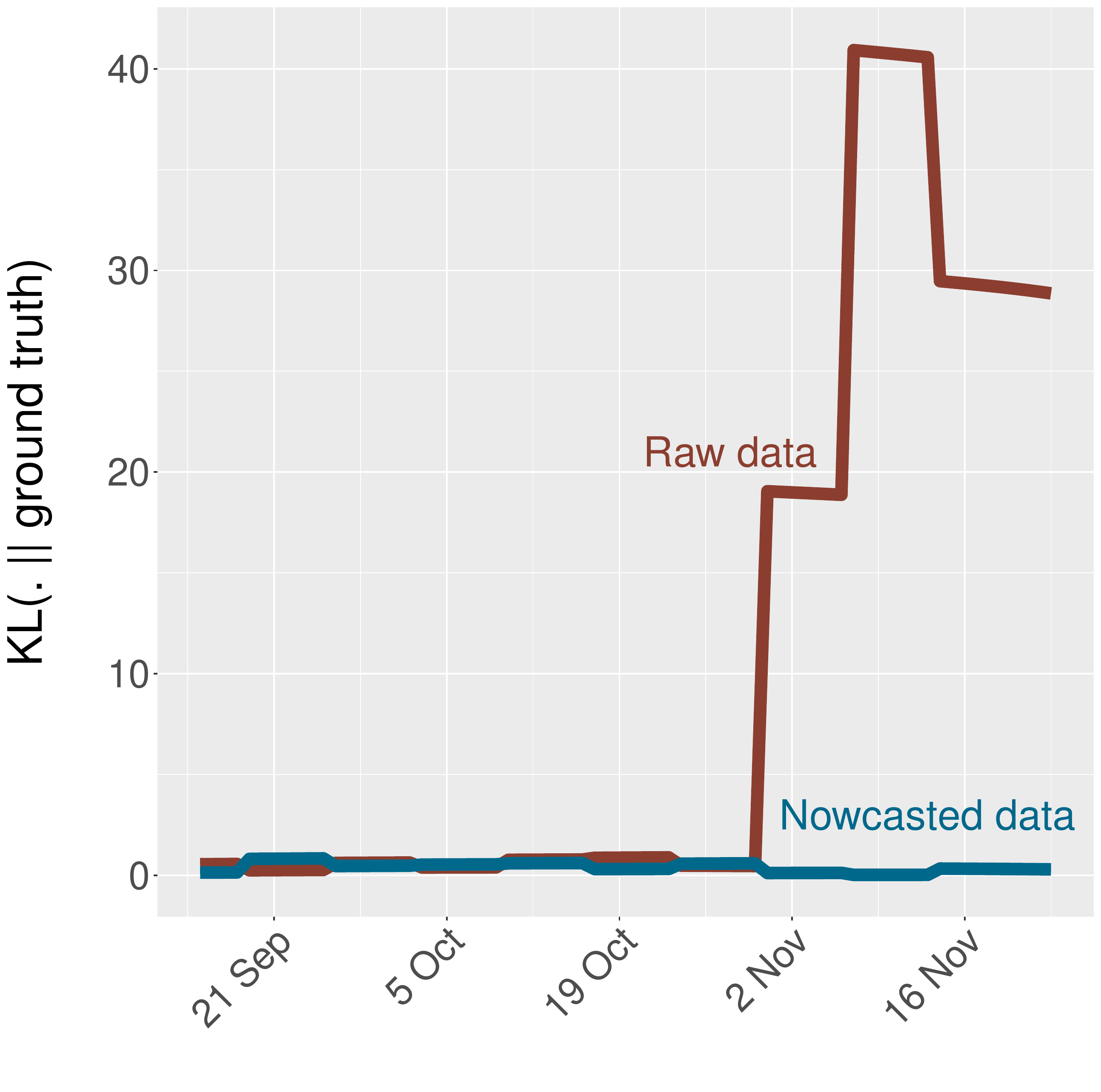}
\par\end{centering}
\caption{Kullback-Leibler divergence (relative entropy) between the $R_t$ value estimated using raw data and nowcasted data. \label{fig: KL_divergence}}
\end{figure}

Applying nowcasting to surveillance data suffering from the reporting delays is crucial to accurate tracking of real-time epidemic dynamics.
The limitations associated with using non-corrected data in epidemiological analyses is highlighted with our results of the $R_t$ estimates shown in Figure~\ref{fig: Rt}. Use of this raw data leads to continued underestimation of $R_t$ and predicts a declining epidemic. Specifically, in the month preceding the nowcast, the relative entropy value for the ground truth and raw data $R_t$ was on average 13.14 (max 43.8) and for ground truth and nowcasted data $R_t$ only 0.26 (max 0.35) (see Figure~\ref{fig: KL_divergence}). By contrast, the ground truth results show that the epidemic remains uncontrolled, with $R_t$ remaining above 1 - an important conclusion also captured by our nowcasting approach.

\begin{figure}[!h]
\begin{centering}
\includegraphics[scale=0.5]{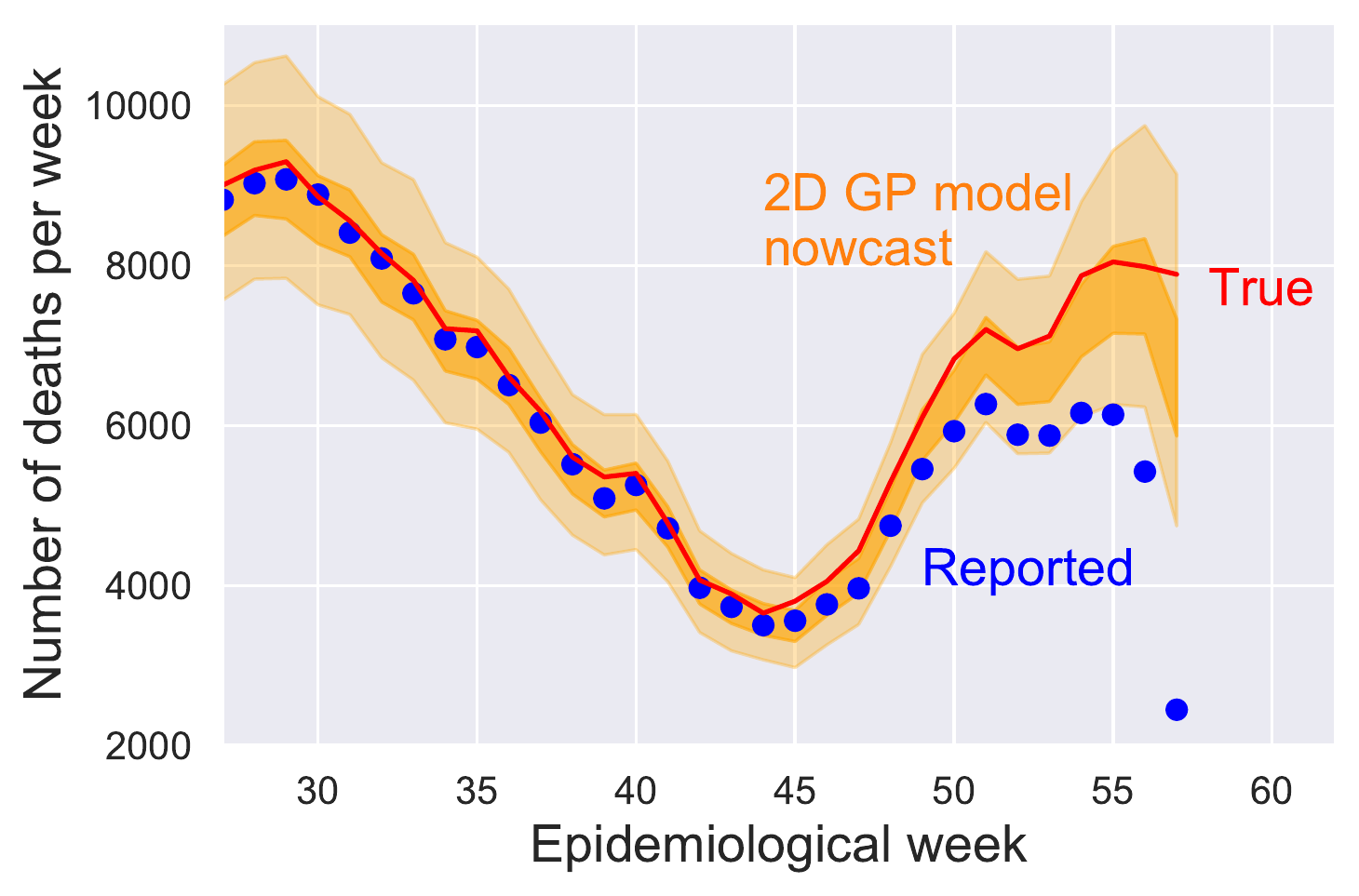}
\par\end{centering}
\caption{Nowcasted and reported deaths due to COVID-19 death for Brazil up to 8-Feb-2021 generated with the 2D additive GP model. 50\% and 95\% CrI for the GP model nowcasts are shown with the ribbon. True values were obtained from the SIVEP release on the 31-May-2021. \label{fig: nowcast_Brazil}}
\end{figure}

The recent emergence of SAR-CoV-2 variants of concern with altered epidemiological characteristics, such as increased transmissibility \citep{volz_nature} or partial evasion of immunity \citep{faria_science}, emphasise the need accurate and continued real-time epidemic tracking.
To estimate COVID-19 mortality in Brazil, during a resurgent phase of the epidemic concurrent with the emergence of the P.1 variant, we perform nowcasting using data released on the 8-Feb-2021 and compare the predictions to the updated numbers of deaths released on the 31-May-2021. The results of the nowcast for the whole of Brazil are shown in Figure~\ref{fig: nowcast_Brazil}. The reported data show a decline in the weekly number of deaths since epidemiological week 54 for Brazil, however the nowcasted results show much higher numbers of estimated weekly deaths, closer to the true value.

\begin{figure*}[!h]
\begin{centering}
\includegraphics[scale=0.5]{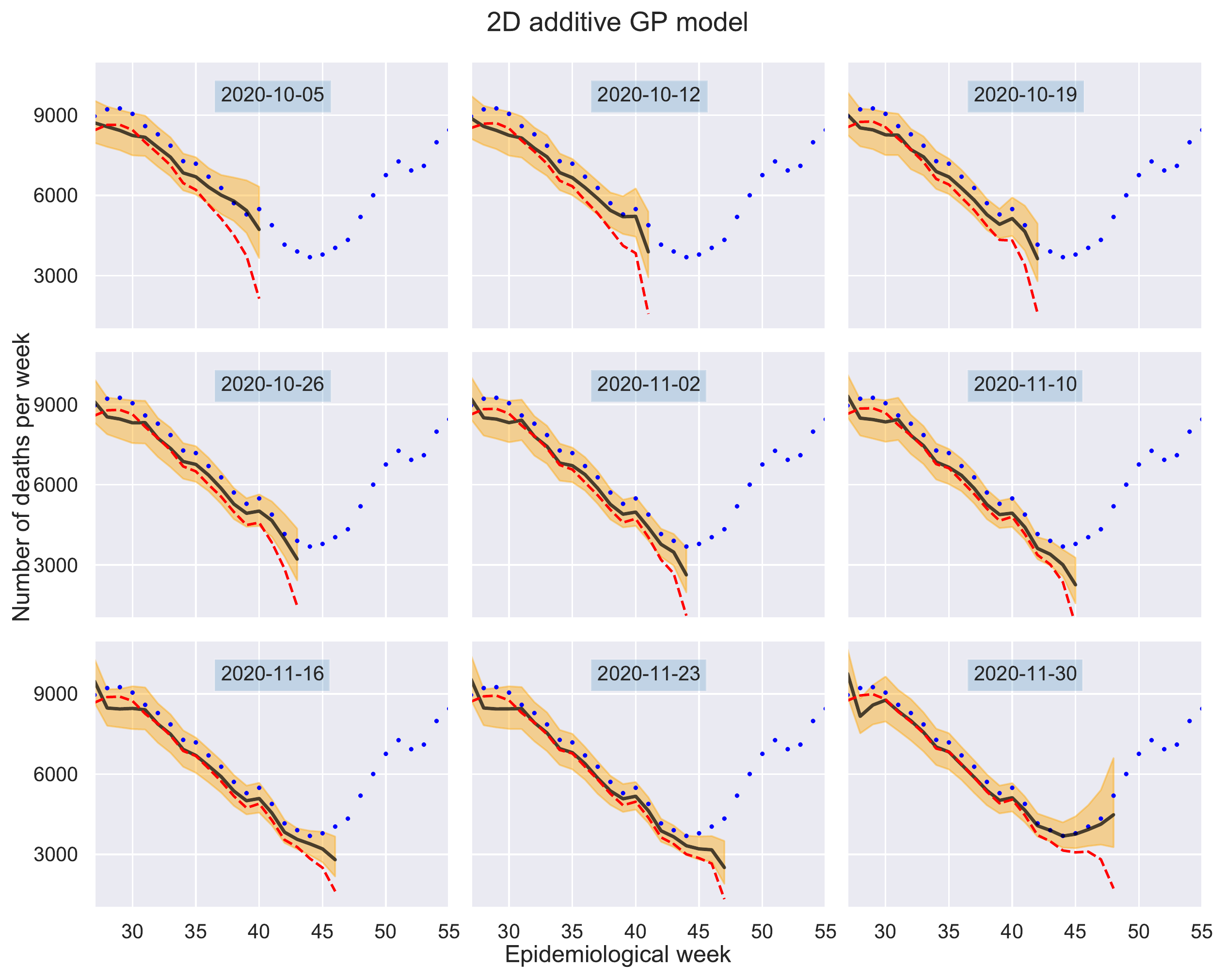}
\par\end{centering}
\caption{Retrospective tests for the 2D additive kernel GP model. \label{fig: all_dates_9}}
\end{figure*}

\begin{figure*}[!h]
\begin{centering}
\includegraphics[scale=0.5]{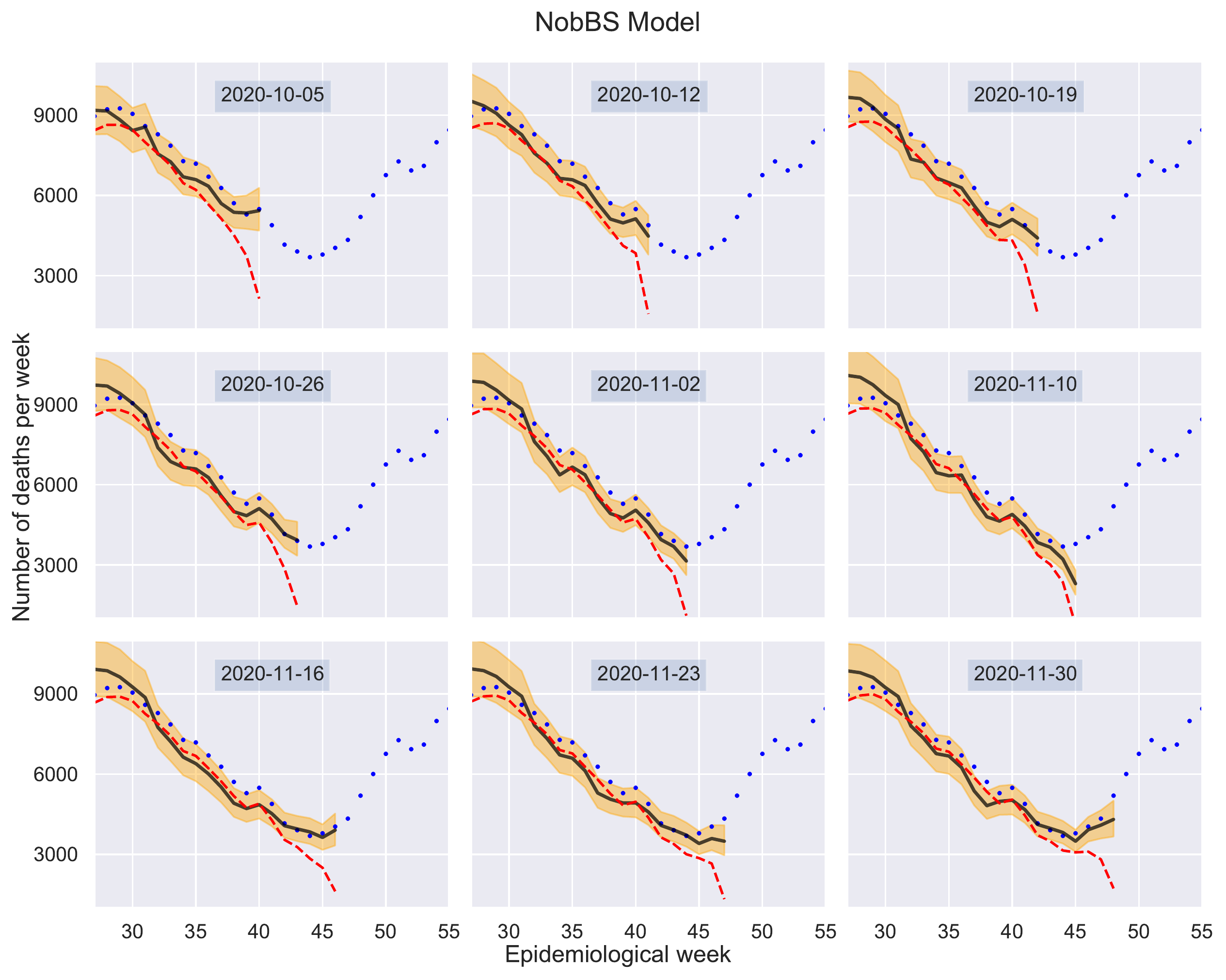}
\par\end{centering}
\caption{Retrospective tests for the NobBS model. \label{fig: all_dates_10}}
\end{figure*}

The GP nowcasting framework we introduced was benchmarked with the established NobBS method \citep{mcgough2020nowcasting}. The main theoretical difference between the two proposed models is how the mean of the negative-binomial distribution, describing the number of deaths per week, is modelled. Our model uses a latent Gaussian process to do that, whereas NobBS uses first-order random walk. The 2D GP model improves on the RMSE for point predictions and CRPS for the distribution of samples compared to NobBS, as shown in Figure~\ref{fig: RMSE_weighted}, 
and provides more realistic uncertainty intervals (Figures~\ref{fig: all_dates_9} and \ref{fig: all_dates_10}). As well as improving predictive performance on the missing parts of the reporting triangle, the GP framework provides a more expressive statistical model capable of better explaining the historical reporting data (see Figure~\ref{fig: nowcast_brazil_delays}).


One of the limitations of the approach described here is the dependence on the historical data and the regularity of the data releases, a limitation shared by many other nowcasting approaches. Additional challenges include variability in the distribution of reporting delays over time. For example, during the initial phase of the Brazilian COVID-19 epidemic, reporting delays were particularly severe. Delays in reporting are typically most extensive during outbreaks of a novel pathogen (such as SARS-CoV-2), due to the limitations in diagnostic availability and testing capacity. Relatedly, during epidemic peaks, strain on healthcare systems and administrative staff due to increasing admissions can also result in lengthening of reporting delays.

\begin{figure*}[!h]
\begin{centering}
\includegraphics[scale=0.55]{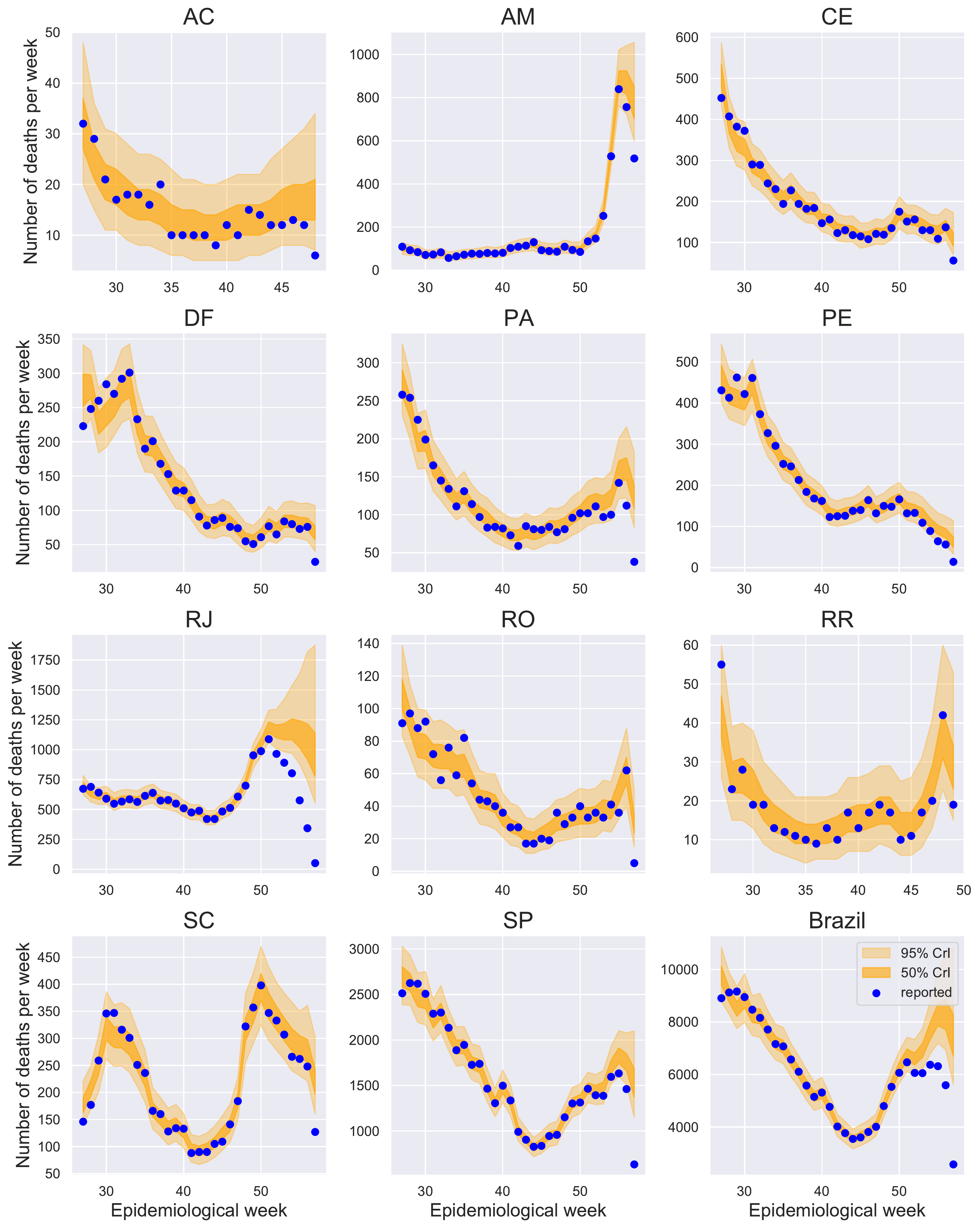}
\par\end{centering}
\caption{Nowcasts made by the 1D SE+SE data-split GP model, using data up to 8-Feb-2021 for Acre (AC), Amazonas (AM), Ceará (CE), Distrito Federal (DF), Pará (PA), Pernambuco (PE), Rio de Janeiro (RJ), Rondônia (RO), Roraima (RR), Santa Catarina (SC), São Paulo (SP) and whole Brazil. \label{fig: many_states}}
\end{figure*}

The GP nowcasting models introduced in this paper can be readily used for real-time monitoring of the new outbreaks of diseases, as relatively few data points are required to train the model, ca. 3 months here. 
In other applications it is possible more data may be required, depending on the distribution of reporting delays, variance of counts, and regularity and granularity of data. Although this paper focuses on application of the proposed GP-based nowcasting framework to the death counts for the whole country, the GP models can be applied at finer spatial scales, as illustrated for individual states of Brazil in Figure~\ref{fig: many_states}. This flexibility is important due to the large heterogeneity of the healthcare system in the country \citep{sivep-ethnic}.

\section{Conclusions}\label{sec:Conclusions}

We have presented a new approach to modelling time-delay data, which can be used to nowcast online data streams that have statistically distributed delays. Our approach uses latent Gaussian processes with additive kernels, and gives a fully flexible and generic method to describe and predict the data for unknown delays. The method has been demonstrated for assessing mortality and estimating the effective reproduction number for COVID-19 reporting in Brazil, but can be used for other contexts in which delays in a measurement process exist.

\section{Code and data availability}
Python, R and Stan code used to analyse the data and fit the nowcasting models is available at \url{https://github.com/ihawryluk/GP_nowcasting}. The SIVEP-Gripe database \citep{SIVEP} is available to download from Brazil Ministry of Health website \url{https://opendatasus.saude.gov.br/dataset/bd-srag-2020}.

\begin{acknowledgements}
The authors thank the CADDE group for insight into COVID-19 reporting in Brazil, and are grateful to Bruce Nelson for consistent public reporting of COVID-19 in Amazonas. T.A.M is grateful to Daniel A. M. Villela for helpful discussion of nowcasting.
The authors would also like to thank the group of anonymous epidemiology experts from Imperial College London Department of Infectious Disease Epidemiology, who participated in testing the nowcasting model against the human predictions.

The authors acknowledge funding from the MRC Centre for Global Infectious Disease Analysis (reference MR/R015600/1), jointly funded by the UK Medical Research Council (MRC) and the UK Foreign, Commonwealth \& Development Office (FCDO), under the MRC/FCDO Concordat agreement and is also part of the EDCTP2 programme supported by the European Union. This research was also partly funded by the Imperial College COVID-19 Research Fund. SB acknowledges The UK Research and Innovation (MR/V038109/1), the Academy of Medical Sciences Springboard Award (SBF004/1080), The MRC (MR/R015600/1), The BMGF (OPP1197730), Imperial College Healthcare NHS Trust- BRC Funding (RDA02), The Novo Nordisk Young Investigator Award (NNF20OC0059309) and The NIHR Health Protection Research Unit in Modelling Methodology. IH was funded by a MRC PhD studentship (MR/S502388/1). CW acknowledges a Medical Research Council Doctoral Training Partnership PhD studentship. 

\end{acknowledgements}

\clearpage
\bibliography{ref.bib}

\newpage
\beginsupplement

\onecolumn
\section{Supplementary Material}

\subsection{Model specification}
\begin{table*}[!h]
\caption{Priors for the analysed models. Here $T$ is the maximum number of weeks for which the data is available, that is rows in the reporting triangle, and $D$ is the maximum reporting delay, that is number of columns in the reporting triangle. $\Gamma$ denotes a Gamma distribution. $^*$ the same hyperparameters were used for the models with Matérn(1/2) and Matérn(3/2) kernels.}
\label{tab: priors}
\begin{tabular}{lllllll}
\toprule
 & \begin{tabular}[c]{@{}c@{}}1D\\SE\\ \end{tabular} &
\begin{tabular}[c]{@{}c@{}}1D\\SE+SE\\ \end{tabular} &
\begin{tabular}[c]{@{}c@{}}1D\\SE+Mat*\\ \end{tabular} &
\begin{tabular}[c]{@{}c@{}}1D SE+SE\\data-split\\ \end{tabular} &
\begin{tabular}[c]{@{}c@{}}2D\\additive\\ \end{tabular} &
NobBS \\
\midrule
$r$ & $\Gamma(500,2)$ & $\Gamma(500,2)$ & $\Gamma(500,2)$ & $\Gamma(500,2)$ & $\Gamma(200,2)$ & $\Gamma(500,2)$ \\
$\alpha_{1,\text{long}}$ & $\mathcal{N}(1,1)$ & $\mathcal{N}(15,2)$ & $\mathcal{N}(15,2)$ & $\mathcal{N}(15,2)$ & - & - \\
$\alpha_{2,\text{long}}$ & - & - & - & $\mathcal{N}(20,2)$ & - & - \\
$\alpha_{1,\text{short}}$ & - & $\mathcal{N}(5,2)$ & $\mathcal{N}(5,1)$ & $\mathcal{N}(5,1)$ & - & - \\
$\alpha_{1,\text{short}}$ & - & - & - & $\mathcal{N}(3,1)$ & - & - \\
$\alpha_{1,t}$ & - & - & - & - & $\mathcal{N}(T,1)$ & - \\
$\alpha_{2,t}$ & - & - & - & - & $\mathcal{N}(D,1)$ & - \\
$\alpha_{1,d}$ & - & - & - & - & $\mathcal{N}(0,1)$ & - \\
$\alpha_{2,d}$ & - & - & - & - & $\mathcal{N}(0,1)$ & - \\
$\rho_{1,\text{long}}$ & $\mathcal{N}(3,1)$ & $\mathcal{N}(T,0.1)$ & $\mathcal{N}(T,0.1)$ & $\mathcal{N}(T,0.1)$ & - & - \\
$\rho_{2,\text{long}}$ & - & - & - & $\mathcal{N}(D,0.1)$ & - & - \\
$\rho_{1,\text{short}}$ & - & $\mathcal{N}(1,0.01)$ & $\mathcal{N}(1,0.01)$ & $\mathcal{N}(1,0.01)$ & - & - \\
$\rho_{2,\text{short}}$ & - & - & - & $\mathcal{N}(1,0.01)$ & - & - \\
$\rho_{1,t}$ & - & - & - & - & $T$ & - \\
$\rho_{2,t}$ & - & - & - & -  & $1$ & - \\
$\rho_{1,d}$ & - & - & - & -  & $D$ & - \\
$\rho_{2,d}$ & - & - & - & -  & $1$ & - \\
$\delta_1$ & $\mathcal{N}(0,1e-6)$ & $\mathcal{N}(0,1e-6)$ & $\mathcal{N}(0,1e-6)$ & $\mathcal{N}(0,1e-6)$ & $\mathcal{N}(0,1e-7)$ & - \\
$\delta_2$ & - & - & - & $\mathcal{N}(0,1e-6)$ & $\mathcal{N}(0,1e-7)$ & - \\
$z$ & $\mathcal{N}(0,0.1)$ & $\mathcal{N}(0,0.1)$ & $\mathcal{N}(0,0.1)$ & $\mathcal{N}(0,0.1)$ & $\mathcal{N}(0,0.1)$ & - \\
$\tau$ & - & - & - & -  & - & $\Gamma(0.01, 0.01)$ \\
$a[1]$ & - & - & - & -  & - & $\mathcal{N}(0, \sqrt{\frac{1}{0.001}})$ \\
$a[t]$ & - & - & - & -  & - & $\mathcal{N}(\alpha[t-1],\sqrt{\frac{1}{\tau}})$ \\
$\beta$ & - & - & - & - & - & $\text{Dirichlet}(0.1)$ \\
\bottomrule
\end{tabular}
\end{table*}

\twocolumn

\begin{figure}[!h]
\begin{centering}
\includegraphics[scale=0.6]{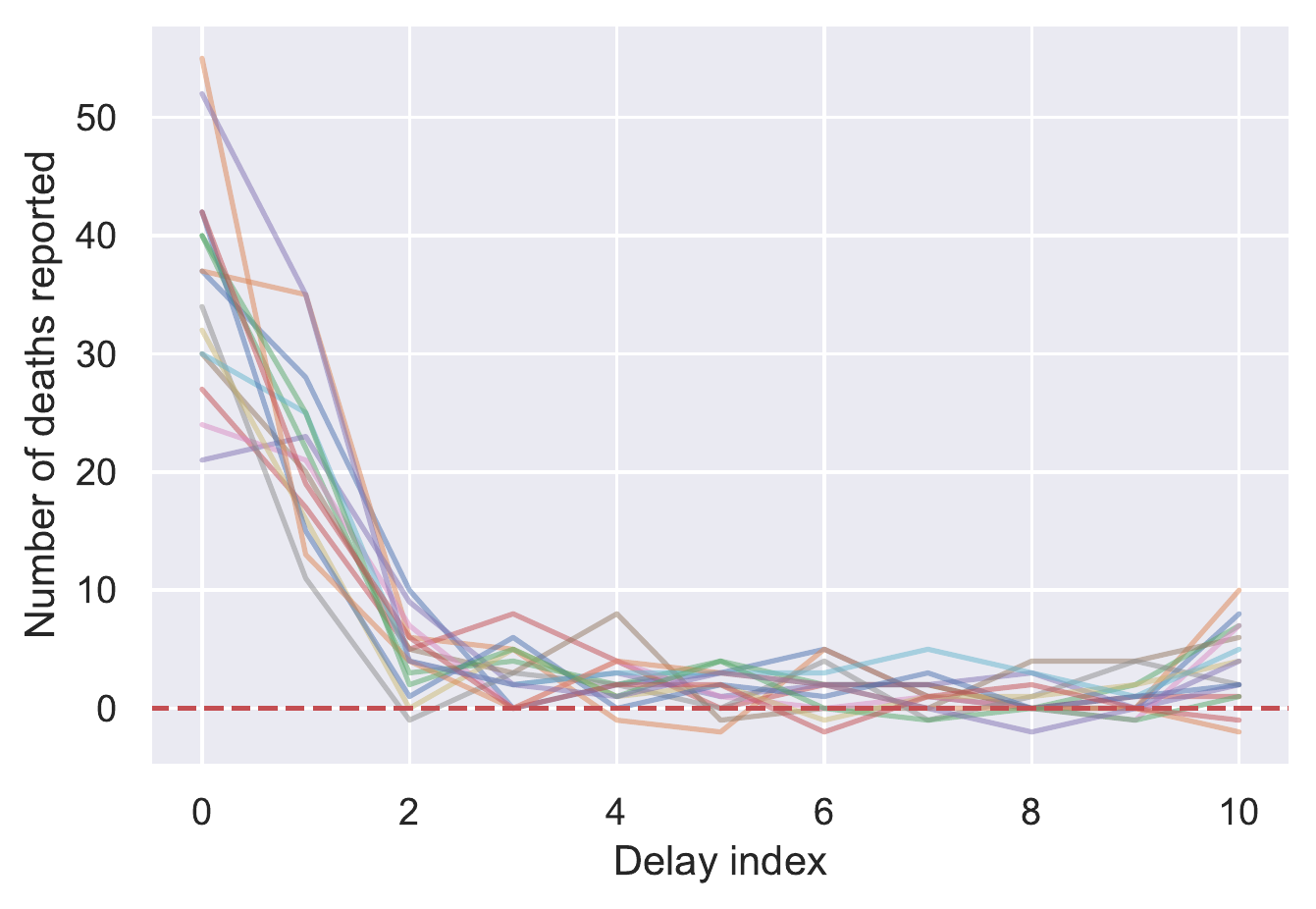}
\par\end{centering}
\caption{Example of weekly reporting delay with negative signal for Amazonas, epidemiological weeks 27 to 42. Each week's mortality data are plotted as a single line. Some lines fall under the $y=0$ line (red dashed line), indicating that there were some deaths that were incorrectly assigned to a given week, which was corrected by the following releases of the data. \label{fig: sivep_errors_AM}}.
\end{figure}

\begin{figure}[!h]
\begin{centering}
\includegraphics[scale=0.5]{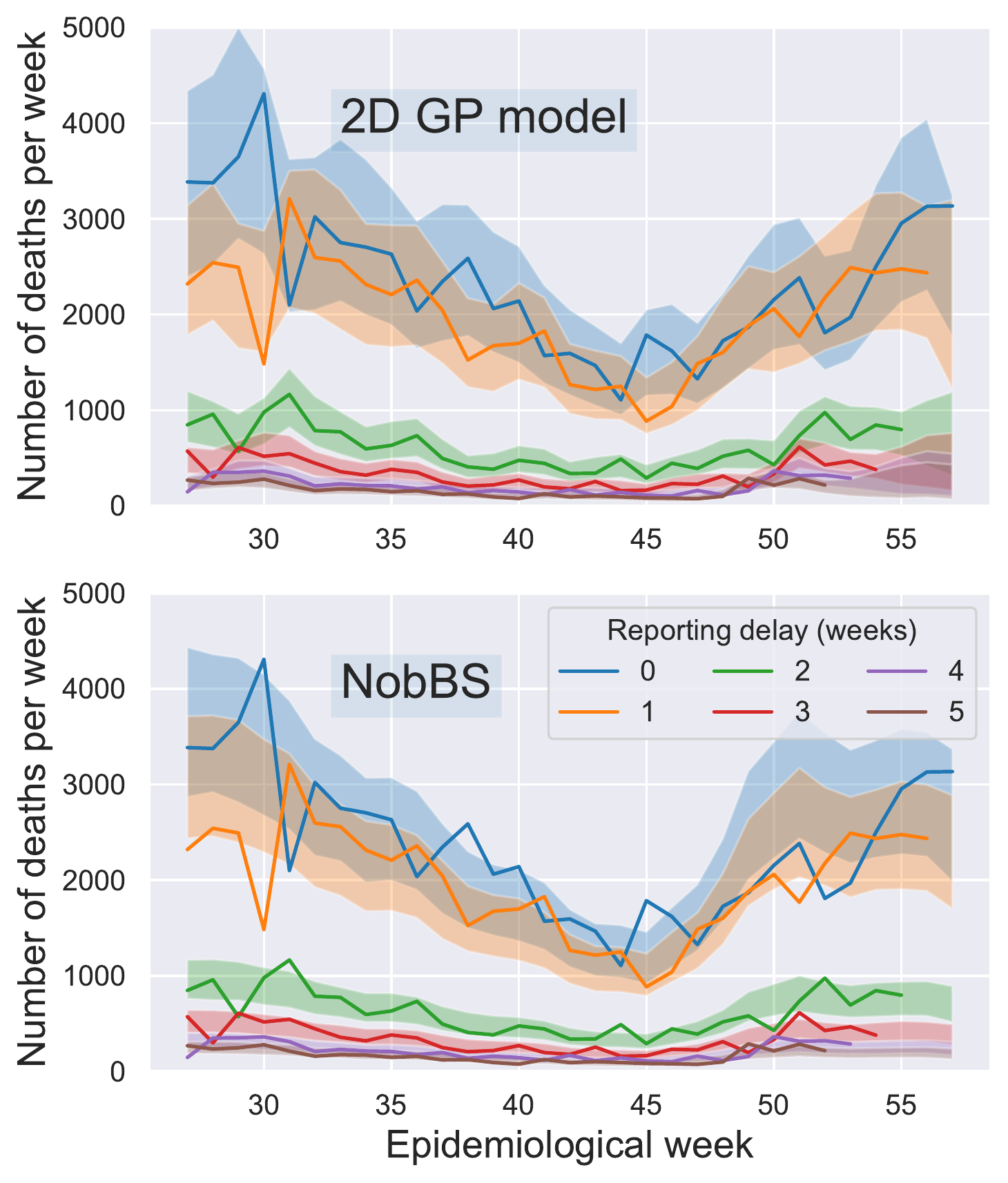}
\par\end{centering}
\caption{Reported and nowcasted numbers of deaths with reporting delay 0 to 5 weeks generated by the 2D additive GP and NobBS models. The reported data are shown with solid lines, and the 95\% CrI for the nowcasts with the ribbons. \label{fig: nowcast_brazil_delays_95crI}}
\end{figure}

\onecolumn

\clearpage
\onecolumn
\subsection{Retrospective testing}

\begin{figure}[!h]
\begin{centering}
\includegraphics[scale=0.3]{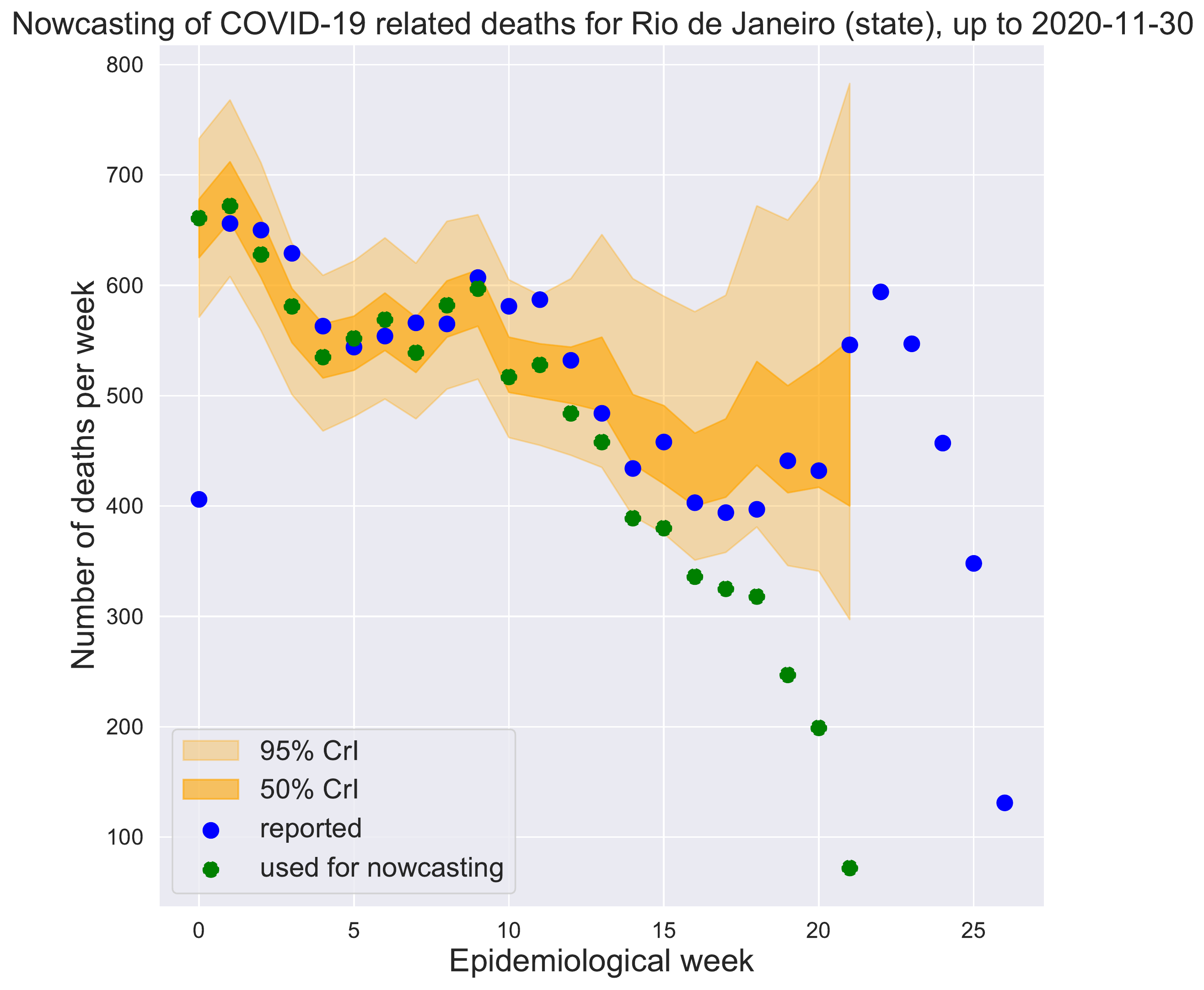}
\par\end{centering}
\caption{Nowcasted and reported deaths due to COVID-19 death for Rio de Janeiro (state), generated with a 1D SE+SE data-split GP model. Reported deaths are shown in blue, nowcasted CrI in orange. Here the nowcasting was performed with all data available up till the SIVEP data release on the 30-Nov-2020. At that point, looking only at the reported data might indicate that the number of deaths keep decreasing, however using nowcasting would have revealed the uptick in the number of deaths, which was not yet observed in the data at the time of the 30-Nov-2020 release. \label{fig: RJ_backtest}}
\end{figure}

\begin{figure*}[!hb]
\begin{centering}
\includegraphics[scale=0.45]{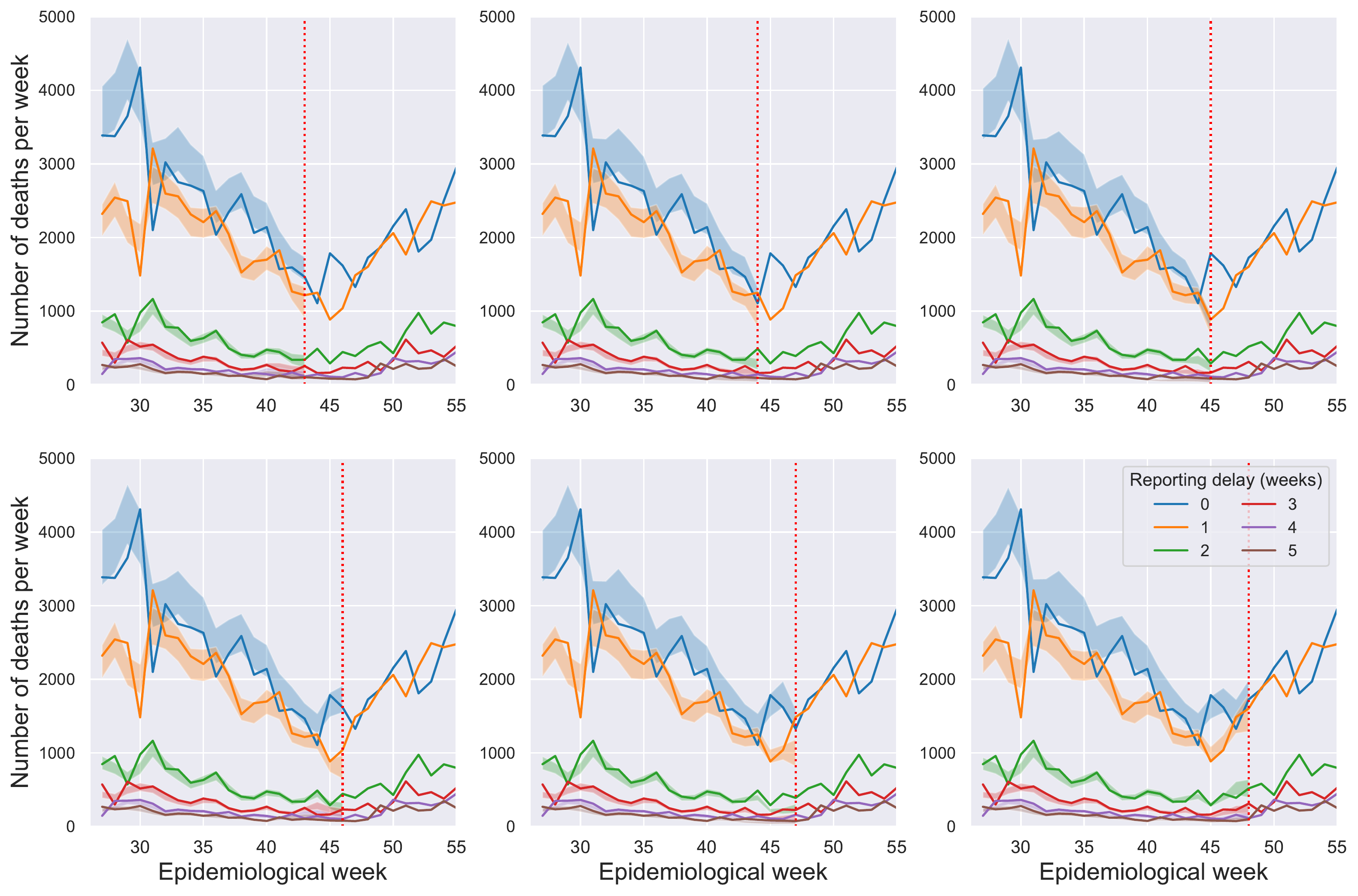}
\par\end{centering}
\caption{Distribution of reporting delays for each of the retrospective tests. The moment of the "nowcast" is shown with the red dotted line in each plot. Solid line present the data extracted from the release of SIVEP data from 31-May-2021, and the ribbons show 50\% CrI of the model fit obtained obtain using the 2D additive kernel GP model. \label{fig: SM_backtest_delays}}
\end{figure*}

\clearpage

\clearpage
\onecolumn
\subsection{Model diagnostics}\label{s: diagnostics}

For each of the model runs shown in this section, 4 chains were run for 1000 iterations, with 500 iterations used for burn-in. All fits presented are done for nowcasting using all data available up to 11-Jan-2021.

\begin{table}[!h]
\caption{Diagnostics for the 1D SE+SE data-split GP model. \label{tab: diagnostics_4compSE}}
\begin{tabular}{lrrrrrrrrr}
\toprule
{} &     mean &      sd &   hdi\_3\% &  hdi\_97\% &  mcse\_mean &  mcse\_sd &  ess\_bulk &  ess\_tail &  r\_hat \\
\midrule
$\rho_{1, \text{long}}$  &   28.004 &   0.101 &   27.820 &   28.194 &      0.002 &    0.001 &    2794.0 &    1241.0 &   1.00 \\
$\rho_{2, \text{long}}$  &   29.009 &   0.102 &   28.826 &   29.209 &      0.002 &    0.002 &    2205.0 &    1377.0 &   1.01 \\
$\rho_{1, \text{short}}$ &    1.003 &   0.010 &    0.984 &    1.021 &      $<$0.001 &    $<$0.001 &    2985.0 &    1388.0 &   1.00 \\
$\rho_{2, \text{short}}$ &    1.013 &   0.009 &    0.996 &    1.032 &      $<$0.001 &    $<$0.001 &    2096.0 &    1691.0 &   1.00 \\
$\alpha_{1, \text{long}}$   &   19.165 &   1.570 &   16.248 &   22.085 &      0.074 &    0.053 &     468.0 &     482.0 &   1.01 \\
$\alpha_{2, \text{long}}$    &   23.366 &   1.234 &   21.145 &   25.692 &      0.084 &    0.060 &     219.0 &     288.0 &   1.02 \\
$\alpha_{1, \text{short}}$   &    4.958 &   0.481 &    4.048 &    5.830 &      0.024 &    0.017 &     412.0 &     460.0 &   1.01 \\
$\alpha_{2, \text{short}}$   &    5.941 &   0.279 &    5.447 &    6.463 &      0.014 &    0.010 &     378.0 &     886.0 &   1.01 \\
$\delta_1$      &    $<$0.001 &   $<$0.001 &    $<$0.001 &    $<$0.001 &      $<$0.001 &    $<$0.001 &    1448.0 &     795.0 &   1.00 \\
$\delta_2$      &    $<$0.001 &   $<$0.001 &    $<$0.001 &    $<$0.001 &      $<$0.001 &    $<$0.001 &     583.0 &    1188.0 &   1.01 \\
$r$           &  246.499 &  11.073 &  225.669 &  267.226 &      0.217 &    0.154 &    2632.0 &    1399.0 &   1.00 \\
\bottomrule
\end{tabular}
\end{table}

\begin{table}[!h]
\caption{Diagnostics for the 2D additive GP model. \label{tab: diagnostics_2Dkron}}
\begin{tabular}{lrrrrrrrrr}
\toprule
{} &    mean &     sd &  hdi\_3\% &  hdi\_97\% &  mcse\_mean &  mcse\_sd &  ess\_bulk &  ess\_tail &  r\_hat \\
\midrule
$\alpha_{1,t}$ &  29.169 &  1.015 &  27.467 &   31.164 &      0.018 &    0.013 &    3288.0 &    1544.0 &   1.01 \\
$\alpha_{2,t}$ &   5.067 &  0.942 &   3.360 &    6.824 &      0.043 &    0.031 &     469.0 &     764.0 &   1.01 \\
$\alpha_{1,d}$ &   0.548 &  0.119 &   0.346 &    0.778 &      0.005 &    0.004 &     471.0 &     680.0 &   1.01 \\
$\alpha_{2,d}$ &   0.616 &  0.096 &   0.460 &    0.805 &      0.004 &    0.003 &     654.0 &     914.0 &   1.01 \\
$\delta_1$       &   $<$0.001 &  $<$0.001 &   $<$0.001 &    $<$0.001 &      0.000 &    $<$0.001 &    1776.0 &     854.0 &   1.00 \\
$\delta_2$      &   $<$0.001 &  $<$0.001 &   $<$0.001 &    $<$0.001 &      $<$0.0010 &    $<$0.001 &    1672.0 &    1110.0 &   1.01 \\
$r$            &  89.733 &  6.478 &  77.809 &  102.541 &      0.176 &    0.126 &    1390.0 &    1365.0 &   1.00 \\
\bottomrule
\end{tabular}
\end{table}

\begin{figure*}[!h]
\begin{centering}
\includegraphics[scale=0.4]{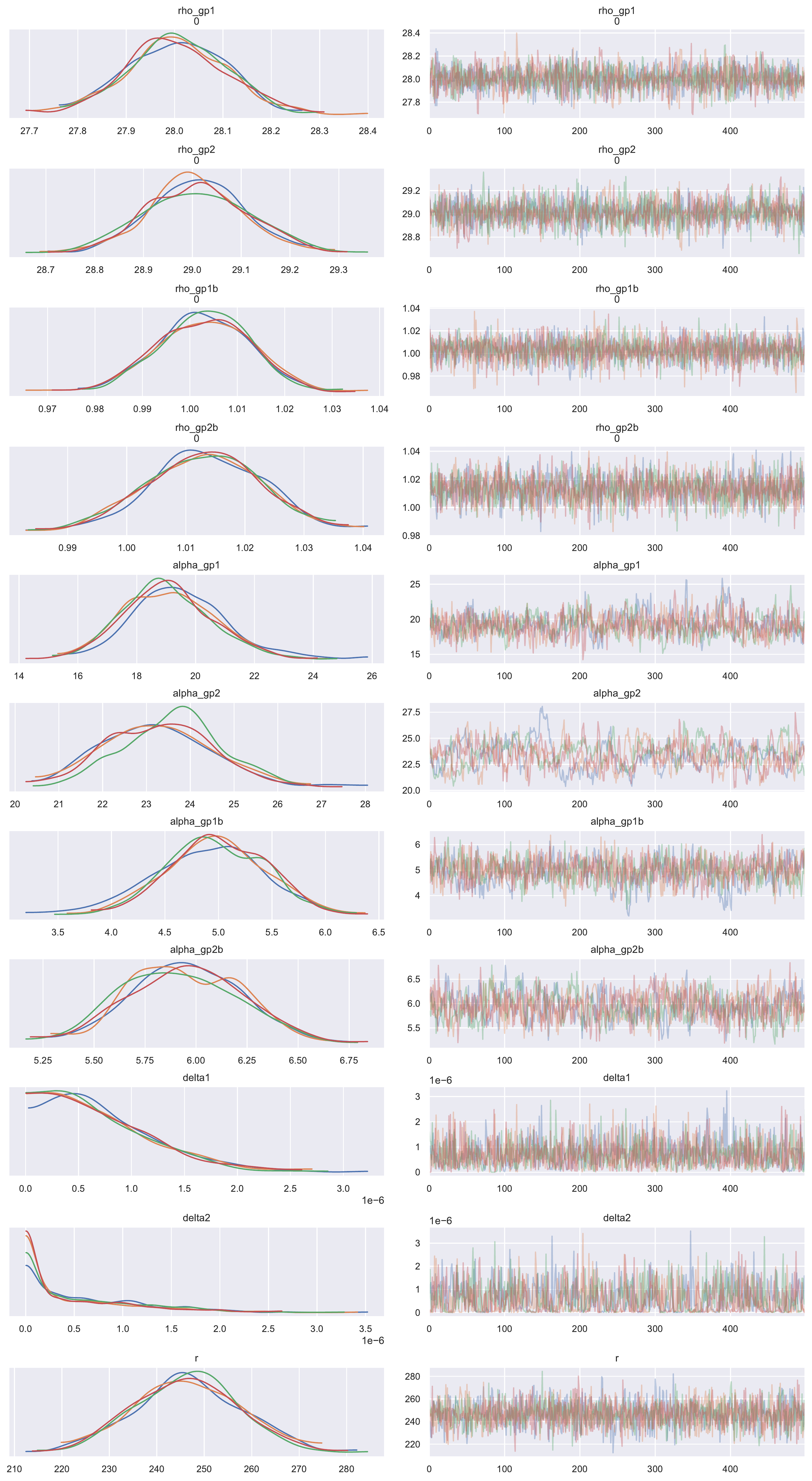}
\par\end{centering}
\caption{Traceplots for the 1D SE+SE data-split GP model.  \label{fig: trace_4compSE}}
\end{figure*}

\begin{figure*}[!h]
\begin{centering}
\includegraphics[scale=0.4]{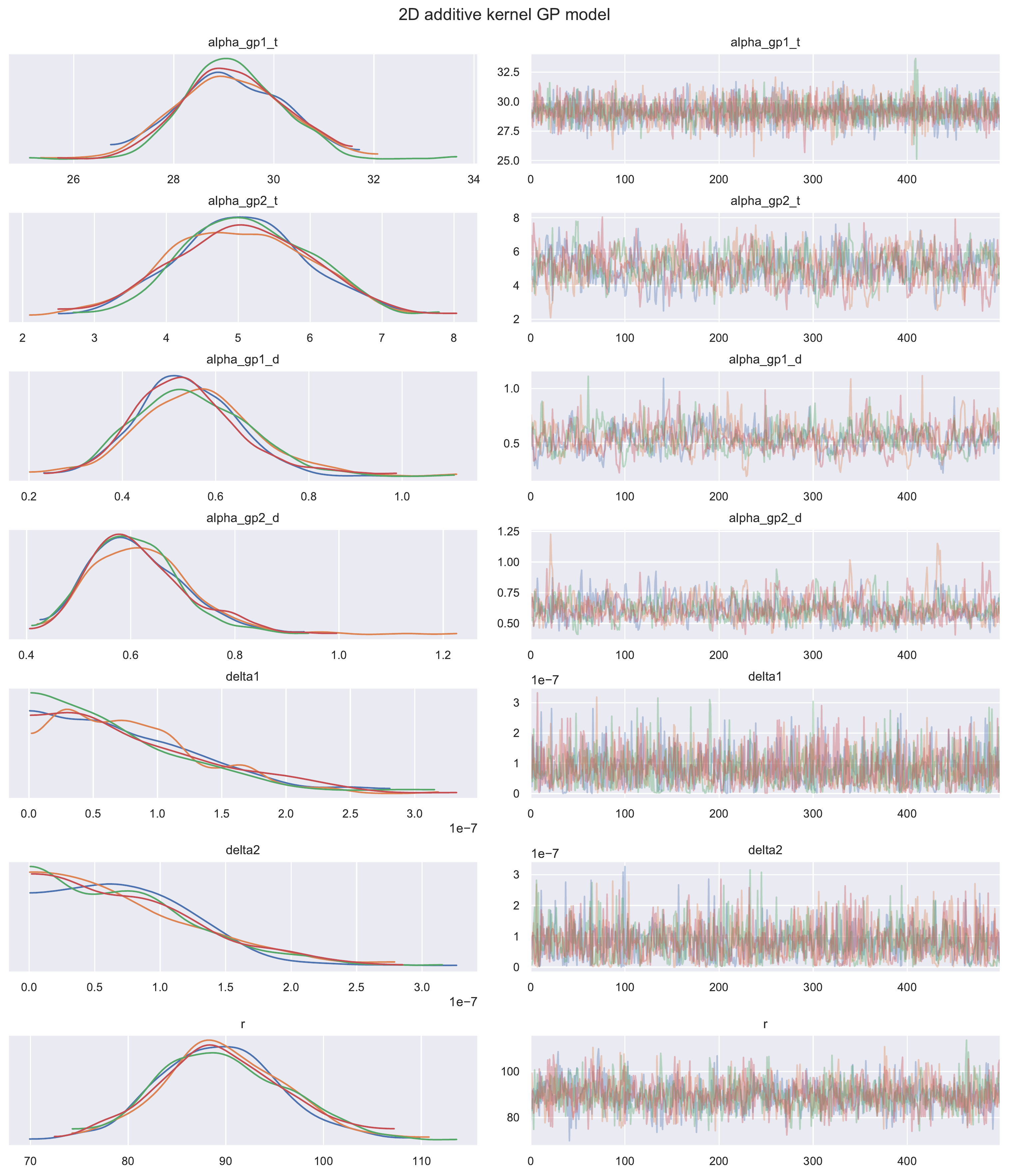}
\par\end{centering}
\caption{Traceplots for the 2D additive GP model. \label{fig: trace_2Dkron}}
\end{figure*}

\clearpage
\subsection{Sensitivity analysis}
For all sensitivity analyses, 1D SE+SE data-split GP model has been used. Each time we run the models with 4 chains for 1000 iterations, with 400 iterations used for burn-in. All fits presented are done for nowcasting using all data available up to 11-Jan-2021.

\end{document}